\documentclass[
backend=bibtex,
reprint,
superscriptaddress,
preprintnumbers,
nobibnotes,nofootinbib,
amsmath,amssymb,
aps,
floatfix,
]{revtex4-1}

\usepackage[normalem]{ulem}
\usepackage{gensymb}
\usepackage{upgreek}
\usepackage{graphicx}
\usepackage{dcolumn}
\usepackage{bm}
\usepackage{floatrow}
\usepackage{float}
\usepackage{hyperref}
\usepackage{xcolor}
\usepackage{textcomp}
\usepackage{amsmath}
\usepackage{varwidth}
\usepackage{gensymb}
\hypersetup{
    colorlinks,
    linkcolor={red!50!black},
    citecolor={blue!50!black},
    urlcolor={blue!80!black}
}
\usepackage[center]{titlesec}
  \titleformat{\section}
   {\centering\normalfont\fontsize{10}{10}\bfseries}{\thesection}{1em}{}
  \titlespacing{\section}{0pt}{12pt plus 4pt minus 2pt}{6pt plus 2pt minus 2pt}
    \titleformat{\subsection}
   {\centering\normalfont\fontsize{10}{10}\bfseries}{\thesubsection}{1em}{}
  \titlespacing{\subsection}{0pt}{12pt plus 4pt minus 2pt}{6pt plus 2pt minus 2pt}
  
\newcommand{\unm}{Center for High Technology Materials and Department of Physics and Astronomy, University of New Mexico, Albuquerque, NM, USA}

\newcommand{\google}{Google Inc., Mountain View, CA, USA}

\begin{document}

    \title{Achromatic varifocal metalens for the visible spectrum}
    
    \author{Maxwell D. Aiello$^{\mathsection}$}
    \affiliation{\unm}
    
    \author{Adam S. Backer$^{\mathsection}$}
    \email{abacke@sandia.gov}
    \affiliation{\unm}
    \affiliation{Sandia National Labs, Albuquerque, NM, USA}

    \author{Aryeh J. Sapon}
    \affiliation{\unm}
    
    \author{Janis Smits}
    \affiliation{\unm}
    \affiliation{Laser Center of the University of Latvia, Riga, LV-1586, Latvia}
    
    \author{John D. Perreault}
    \affiliation{\google}
    
    \author{Patrick Llull}
    \affiliation{\google}
    
    \author{Victor M. Acosta}
    \email{vmacosta@unm.edu}
    \affiliation{\unm}
    
    \renewcommand{\thefootnote}{}{\footnote{$\mathsection$ M. D. Aiello and A. S. Backer contributed equally to this work.}}
    
    \date{\today}

\begin{abstract}
    Metasurface optics provide an ultra-thin alternative to conventional refractive lenses. A present challenge is in realizing metasurfaces that exhibit tunable optical properties and achromatic behavior across the visible spectrum. Here, we report the design, fabrication, and characterization of metasurface lenses (``metalenses'') that use asymmetric TiO$_2$ nanostructures to induce a polarization-dependent optical response. By rotating the polarization of linearly-polarized input light, the focal length of a $40\mbox{-}{\rm \upmu m}\mbox{-}$diameter metalens is tuned from $220\mbox{--}550~{\rm \upmu m}$ with nearly diffraction-limited performance. We show that imparting a wavelength-dependent polarization rotation on incident light enables achromatic focusing over a wide band of the visible spectrum, 483-620 nm. We use this property to demonstrate varifocal color imaging with white light from a halogen source. Tunable achromatic metalenses may be useful for applications in imaging and display. 
\end{abstract}

\maketitle

%\begin{bibunit}

\section{\label{sec:Introduction}Introduction}
Optical lenses with adjustable focal length (``varifocal'') have numerous applications including automated focusing/zooming in photography, volumetric imaging in microscopy~\cite{Mermillod-Blondin2008High-speedLens,Murali2009Three-dimensionalLens}, and in the development of displays that mimic the eye's natural depth cues~\cite{Lanman2013Near-eyeDisplays, Konrad2017Accommodation-invariantDisplays,Padmanaban2017OptimizingDisplays.}. Conventional varifocal lenses based on compound optical systems with mechanically-adjustable spacings~\cite{JosephM.Geary2015IntroductionDesign} are not well suited for applications requiring slim form factors. Alternative varifocal systems based on tunable acoustic gradient index of refraction lenses~\cite{Mermillod-Blondin2008High-speedLens}, liquid lenses~\cite{Murali2009Three-dimensionalLens}, electrowetting~\cite{Won2014Electrowetting-basedDisplay,Li2016ZoomLenses}, spatial light modulators~\cite{Matsuda2017FocalDisplays}, and deformable mirrors~\cite{Kner2010High-resolutionFocusing} increase focal-adjustment speeds and simplify mechanical actuation. However challenges remain in integrating these components in the ultra-compact format desirable for some display and microscopy applications.

Recently developed metasurface optics \cite{Fattal2010FlatAbilities, Lin2014DielectricElements.,Arbabi2015Subwavelength-thickTransmitarrays,Khorasaninejad2016MetalensesImaging.,Zhan2016Low-ContrastOptics} offer the possibility for sub-micrometer-thick optical components with capabilities rivaling those of conventional refractive optics. Metasurface lenses (``metalenses'') use nanostructured dielectric surfaces to impart a spatially varying phase shift on incident light, mimicking how refractive lenses operate. Numerous studies~\cite{Fattal2013ADisplay,Lin2017OpticalWavelengths,Shi2018Polarization-dependentDisplays,Lin2019AchromaticImaging} have demonstrated the capability of metalenses to solve some of the integration challenges associated with miniature optical systems. However, development of achromatic varifocal metalenses operating in the visible spectrum (400-700 nm) remains an ongoing challenge.

One strategy to realizing varifocal metalenses uses electromechanical actuation to strain individual metasurfaces~\cite{Kamali2016HighlyLenses} or modify the lateral displacement of composite ``Alvarez'' metasurfaces~\cite{Zhan2017MetasurfaceNanophotonics,Colburn2018VarifocalMetalenses}. Such metalenses offer broadly tunable focal lengths ($\gtrsim 60\%$~\cite{Kamali2016HighlyLenses}) and even the ability to correct for some imaging aberrations~\cite{She2018AdaptiveShift}. However, with this approach the morphology of the metasurface is altered, rendering it difficult to realize achromatic performance.

Another strategy uses birefringent metasurfaces~\cite{Arbabi2015DielectricTransmission,BalthasarMueller2017MetasurfacePolarization} to tune optical function by varying the polarization of incident light. With this approach, polarization-dependent bi-focal~\cite{Schonbrun2011ReconfigurableNanowires} and bi-chromatic~\cite{Arbabi2016HighMeta-atoms} metalenses have been demonstrated at near-infrared wavelengths. Notable applications of polarization-sensitive metasurfaces include two-photon microscopy \cite{Arbabi2018Two-PhotonLens}, chiral imaging \cite{Khorasaninejad2016MultispectralMetalens}, and orientation-independent single-molecule fluorescence localization imaging \cite{Backlund2016RemovingMask}. However, most implementations of this approach were designed to have exactly two distinct functions (corresponding to orthogonal input polarizations), as opposed to a range of tunability.

In this manuscript, we demonstrate a metalens with a tunable focal length, realized by rotating the polarization of linearly-polarized input light. We show that imparting a wavelength-dependent polarization rotation on incident light enables achromatic focusing over a wide band of the visible spectrum, 483-620 nm. We use this property to demonstrate varifocal color-imaging with white light from a halogen source, using the same metalens. 

\begin{figure}[ht]
    \centering
\includegraphics[width=.92\textwidth]{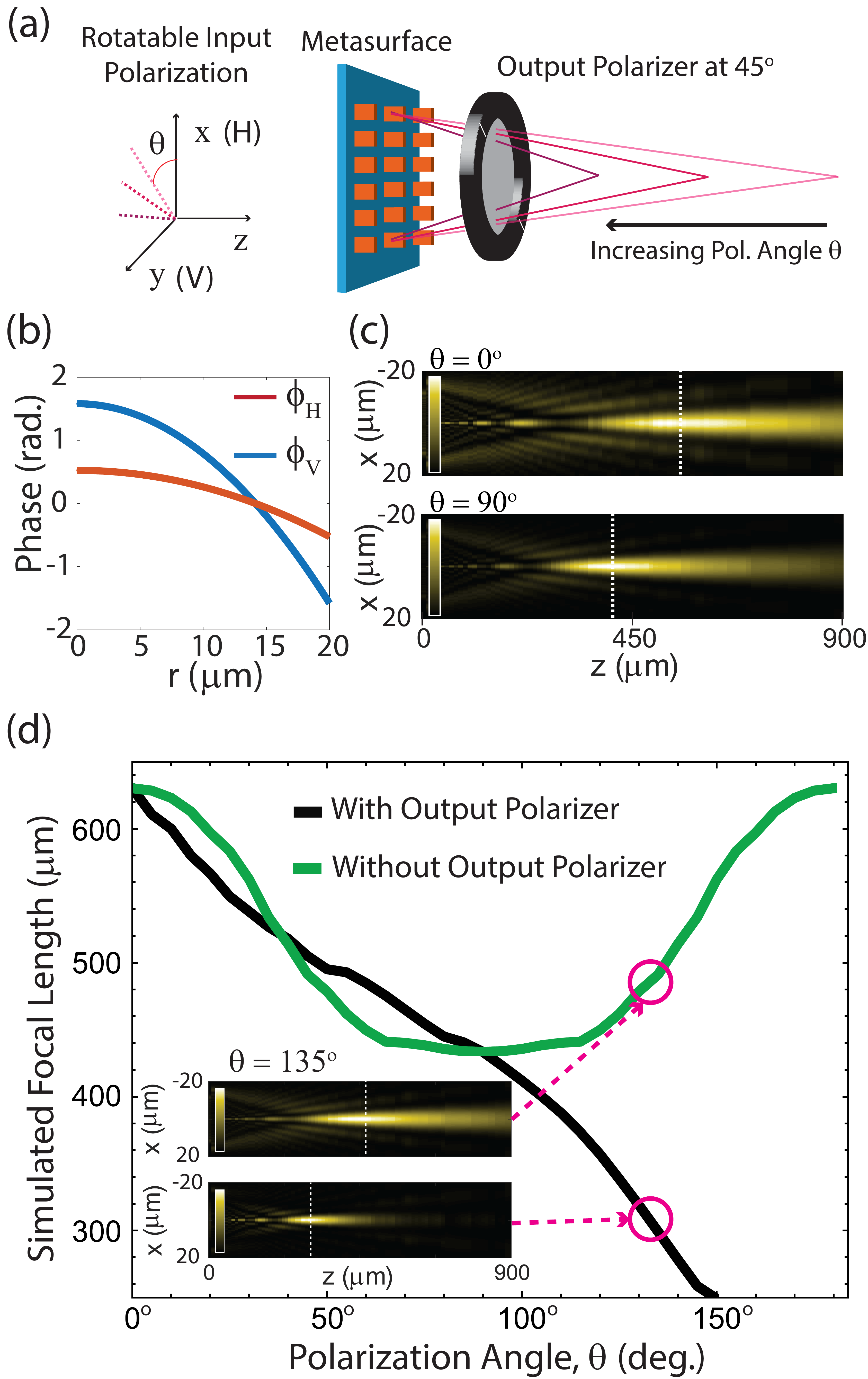}\hfill
       \caption{\label{fig:1}
\textbf{Design of a varifocal metalens.} a) Principle of operation: Linearly polarized light passes through a metalens and an output polarizer oriented at 45$\degree$. The focal length changes as the input polarization is rotated. b) The metalens is designed to have a parabolic phase that varies with radial distance ($r=\sqrt{x^2+y^2}$) from the center of the aperture. The parabolic phase function for vertically polarized light $(\phi_V)$ bends more sharply than that of horizontally polarized light ($\phi_H$). c) Simulated XZ-slices of the focal spot for H- ($\theta = 0\degree$) and V-polarized ($\theta = 90\degree$) light of wavelength $\lambda = 581~{\rm nm}$. The greater phase curvature associated with $\phi_V$ induces a shorter focal length. The fitted focal lengths are annotated as dashed white lines. d)  Simulated metalens focal length ($\lambda = 581~{\rm nm}$) versus input polarization, for the cases with and without the output polarizer. The output polarizer more than doubles the focal tuning range. (Inset) XZ-slices with and without the output polarizer for input polarization $\theta=135\degree$. 
       }
\end{figure}

\section{\label{sec:sim}Principle of operation}
Figure~\ref{fig:1}(a) illustrates the concept of our varifocal metalens design. A birefringent metalens is constructed with different focal lengths for H ($\theta=0\degree$) and V ($\theta=90\degree$) linear polarizations, $f_{\rm H}$ and $f_{\rm V}$ respectively. The metasurface imparts a polarization-dependent, spatially varying phase shift, $\phi_{\rm H,V}(x,y)$, to an incoming wavefront propagating along the $z$ direction. Using the paraxial formula for a thin lens \cite{JosephW.Goodman1996IntroductionOptics}, these phase shifts are given as:

\begin{equation}
\label{eqn:thinlens}
\begin{split}
    \phi_{\rm H}(x,y) & \approx -\alpha_{\rm H}(f_{\rm H}, \lambda)\cdot\left(x^2 + y^2\right) \\     
    \phi_{\rm V}(x,y) & \approx -\alpha_{\rm V}(f_{\rm V}, \lambda)\cdot\left(x^2 + y^2\right) ,
    \end{split}
\end{equation}
where $\lambda$ is the optical wavelength and $\alpha_{\rm H,V}(f_{\rm H,V},\lambda)$ denotes the curvature of the parabolic phase profiles. In the limit $\mathit{NA}\cdot D/\lambda>>1$, where $D$ is the lens diameter and $\mathit{NA}$ is the numerical aperture,  $\alpha_{\rm H,V}(f_{\rm H,V},\lambda)=\pi/(\lambda f_{\rm H,V})$, Sec.~\ref{sec:SI1B}.

Figure~\ref{fig:1}(b) shows the radial phase profiles for a simulated metalens ($40~{\rm \upmu m}$ diameter) with $f_{\rm H}=635~{\rm \upmu m}$ and $f_{\rm V}=435~{\rm \upmu m}$ at $\lambda=581~{\rm nm}$. 
The corresponding cross-sectional intensity profiles (``XZ-slices'') for an input plane wave are also shown. For arbitrary input polarization, the metalens output intensity is a linear combination of H- and V-polarized responses. By continuously varying the polarization angle from $\theta=0\degree$ versus $\theta=90\degree$, the focal point smoothly varies between $f_{\rm H}$ and $f_{\rm V}$, Fig.~\ref{fig:1}(c). Such a continuous response is possible because $f_{\rm H}-f_{\rm V}$ is smaller than the lens depth of field (approximately $200\mbox{--}300~{\rm \upmu m}$). Figure~\ref{fig:1}(d)(top inset) shows the XZ-slice for a polarization angle of $\theta=135\degree$ (or, equivalently, $\theta=45\degree$), where the focal point is near the midpoint between $f_{\rm H}$ and $f_{\rm V}$.

An insight of the present manuscript is that introducing a polarizer after the metalens extends the focal tuning range beyond the limits set by $f_{\rm H}$ and $f_{\rm V}$. The output polarizer projects H and V polarized light onto a single axis at $\theta=45\degree$, causing interference between the H and V electric field components. This interference modifies the output phase of the optical system to a form that continuously varies with $\theta$, unlike the binary case in Eq.~\eqref{eqn:thinlens}, Sec.~\ref{sec:SI1B}. Furthermore, the output polarizer breaks the reflective symmetry about the x- and y-axes of the optical system. This feature causes the optical system to exhibit a different focal length for input light polarized at, for example, $\theta=45\degree$ versus $\theta=135\degree$. Figure~\ref{fig:1}(d) shows the polarization dependence of the focal length when the $45\degree$ output polarizer is inserted after the metalens. The focal tuning range of the simulated  metalens is more than two times wider than the case without the output polarizer. While the output polarizer introduces a polarization-dependent intensity loss, the focusing properties of the system are similar to that of an ideal lens, Fig.~\ref{fig:1}(d)(bottom inset). A detailed derivation of the response of the optical system is found in Sec.~\ref{sec:SI1B}.

\begin{figure}[ht]
    \centering
\includegraphics[width=.98\textwidth]{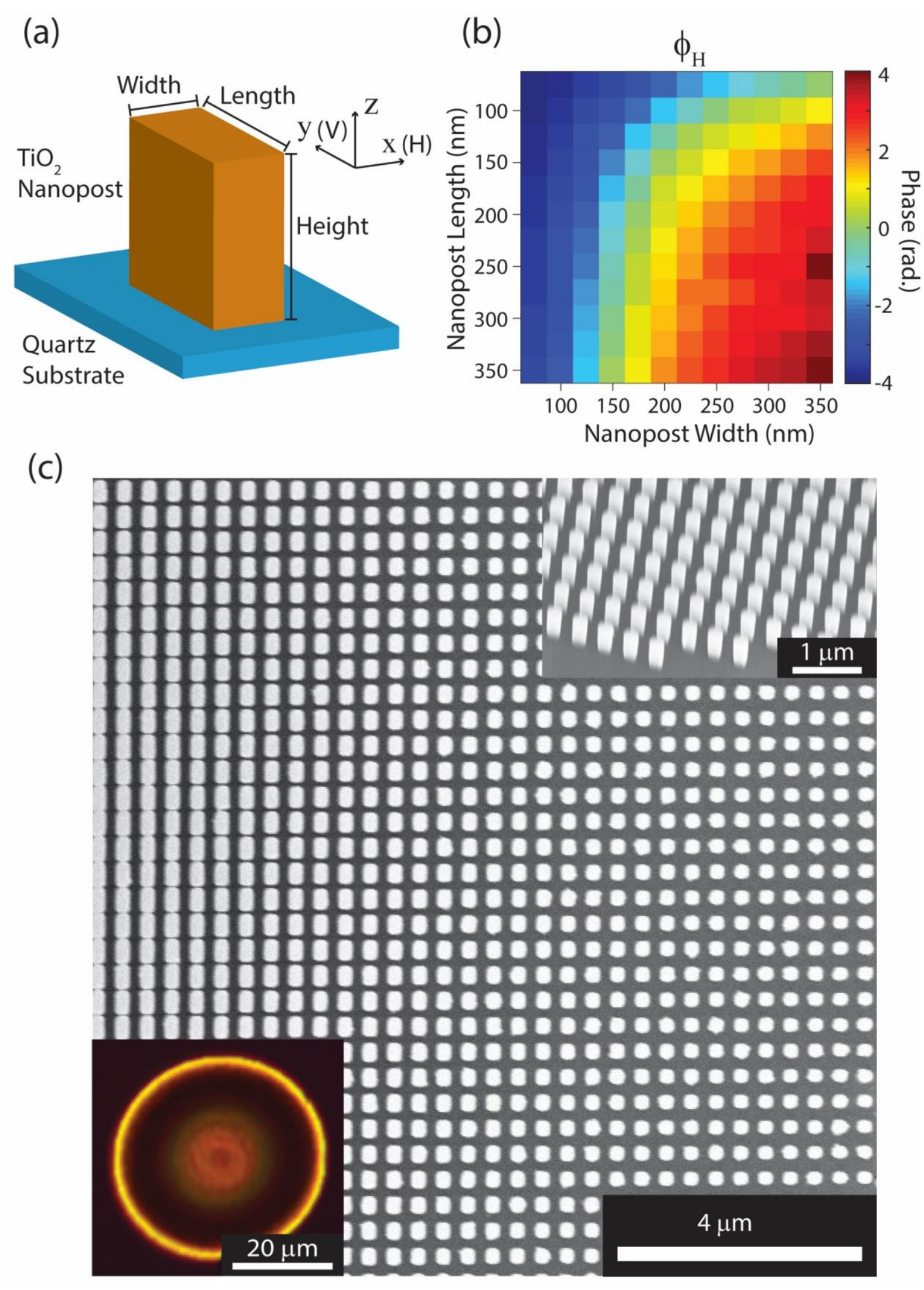}\hfill
       \caption{\label{fig:idea}
\textbf{Simulation and fabrication of TiO$_2$ metalenses.} a) Metalenses consist of an array of titanium-dioxide nanoposts, of variable length and width, on a quartz substrate. The length and width of the nanoposts vary across the diameter of the metalens, while the height and pitch are kept constant at $600~{\rm nm}$ and $400~{\rm nm}$, respectively. b) Relative change in phase for H-polarized light, $\phi_H$, as a function of nanopost length and width. Simulation data is shown for the representative wavelength $\lambda = 581~{\rm nm}$. $\phi_V$ is given by the transpose of this array. c) Scanning electron microscopy (SEM) image of the metalens studied throughout the manuscript. (Top-right inset) 30$\degree$ tilted projection image. (Bottom-left inset) Optical microscope image of the metalens surface viewed through crossed polarizers. 
       }
\end{figure}

\section{\label{sec:desfab}Simulation and fabrication}

Birefringent metalenses were designed and fabricated using previously published techniques~\cite{Arbabi2015DielectricTransmission,BalthasarMueller2017MetasurfacePolarization,Devlin2016BroadbandSpectrum.}. The metalenses were formed from arrays of rectangular titanium dioxide (TiO$_2$) nanoposts with constant height ($600~{\rm nm}$) and pitch ($400~{\rm nm}$),  Fig.~\ref{fig:idea}(a). To achieve a polarization-dependent phase response, Eq.~\eqref{eqn:thinlens}, nanopost lengths and widths were designed to be asymmetric, such that H- and V-polarized light incurred the phase responses shown in Fig.~\ref{fig:1}(b). Finite-difference time-domain simulation software (Lumerical FDTD Solutions) was used to simulate periodic nanopost arrays of different aspect ratios (Sec.~\ref{sec:SI1A}). These parameter sweeps were used to generate lookup-tables [Fig.~\ref{fig:idea}(b)] for determining the desired nanopost dimensions across the metasurface aperture. 

To fabricate TiO$_2$ arrays \cite{Devlin2016BroadbandSpectrum.}, a positive electron-beam resist (ZEP520A) was spin-coated to a thickness of $600~{\rm nm}$ on a quartz substrate. This thickness defined the height of the final nanoposts. Next, electron-beam lithography was used to create vias in the resist, according to the desired nanopost array pattern. The vias were then filled by depositing a $350~{\rm nm}$ layer of TiO$_2$ using atomic layer deposition. The undesired layer of TiO$_2$ on top of the resist was subsequently removed by inductively-coupled plasma etching. The device was soaked in Remover PG at 70\degree C to remove the remaining resist, revealing the final TiO$_2$-on-quartz metasurface. Metalenses of different diameters (up to 1 mm diameter) and phase profiles were fabricated using this method. Figure~\ref{fig:idea}(c) shows scanning electron micrographs and an optical image of the $40\mbox{-}{\rm \upmu m}\mbox{-}$metalens studied here. Additional details on fabrication are presented in Sec.~\ref{sec:SIfab}.   

\begin{figure*}[htbp]
    \centering
\includegraphics[width=.98\textwidth]{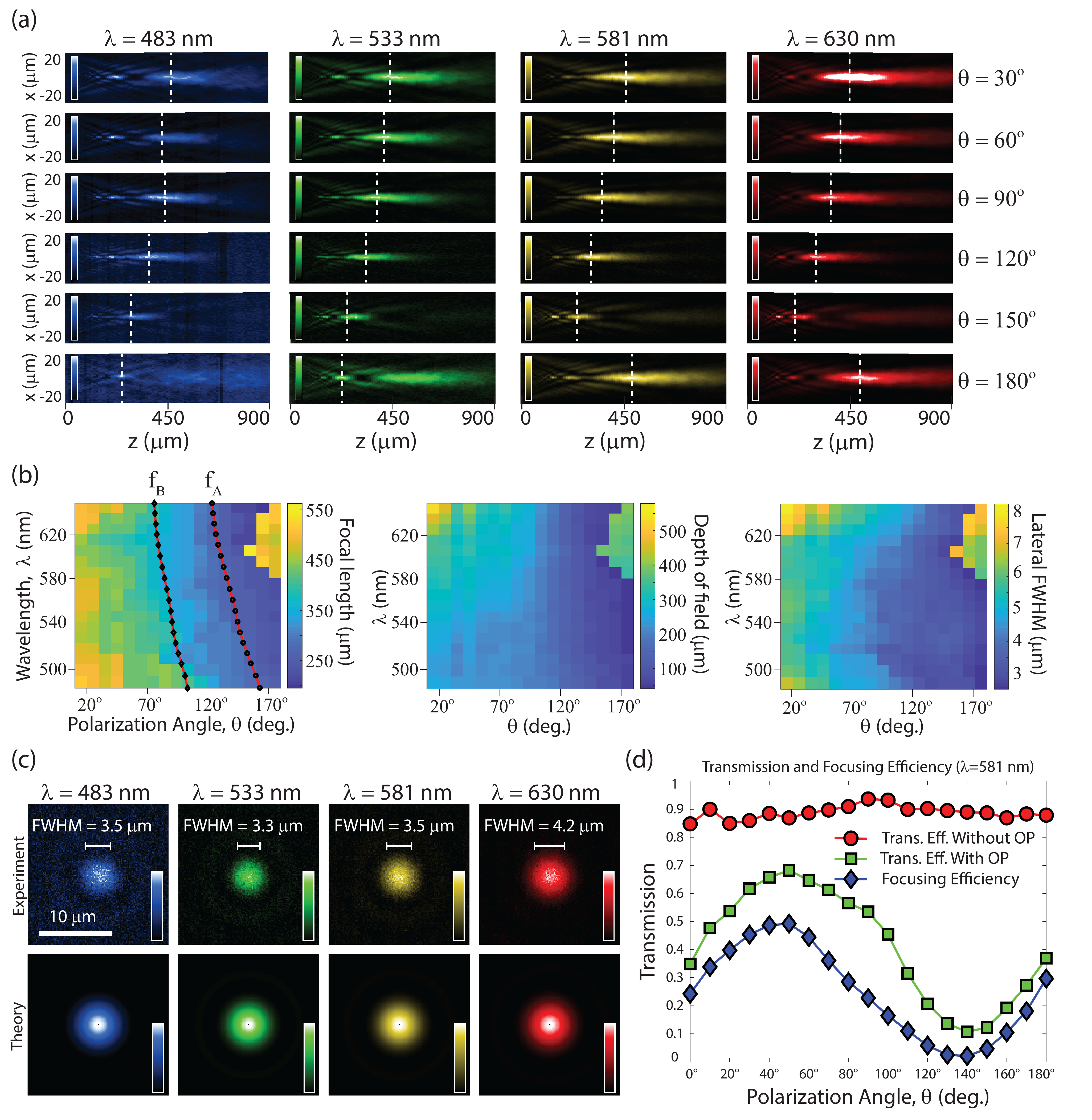}\hfill
       \caption{\label{fig:2}
\textbf{Varifocal metalens characterization.} a) Experimentally-measured XZ-slices of the  metalens PSF. Fitted focal positions are shown as vertical dashed lines. In cases where multiple foci are present, the focal length corresponds to the position with highest intensity. In the few cases where foci had nearly equal intensity, the longer focal length was selected. Intensity scale bars are normalized to largest value in each panel. b) Metalens focal length, depth of field, and lateral FWHM, determined from Gaussian fits to the raw PSF data, as a function of input polarization and wavelength. Dashed black lines refer to two curves of constant focal length, $f_A$\textbf{$=278~{\rm \upmu m}$} and $f_B$\textbf{$=375~{\rm \upmu m}$}. These curves are used for chromatic correction, described later in the text. c) Top: experimentally-measured XY-slices in the focal plane for four representative wavelengths with input polarization $\theta = 120\degree$. The fitted lateral FWHM is annotated on each plot. Bottom: Simulated focal plane intensity distributions, assuming a uniformly transmissive aperture. Intensity scale bars are normalized to largest value in each panel. d) Metalens transmission and focusing efficiency as a function of input polarization, measured at $\lambda = 581~{\rm nm}$. OP--output polarizer. Definitions for these parameters are found in Sec.~\ref{sec:SItransmission}. 
       }
\end{figure*}

\section{\label{subsec:psf}Metalens characterization}
To characterize metalens performance, we measured the three-dimensional intensity output of the combined metalens and output-polarizer system. Using a supercontinuum laser, the metalens was illuminated with a collimated, normally-incident, linearly-polarized light beam covering an area much larger than the metalens aperture. The output intensity profiles comprise the metalens point-spread-function (PSF), which we measured as a function of wavelength $\lambda$ and polarization $\theta$. Details regarding the optical characterization setup, experimental procedure, and data acquisition software are presented in Sec.~\ref{sec:SI3}. 

We measured the three-dimensional (3D) metalens PSF at 19 distinct wavelengths between 483 and 648 nm, and 18 different input polarization angles (10$\degree\mbox{--}180\degree$). Figure~\ref{fig:2}(a) shows XZ-slices of the experimentally-measured PSF at representative  wavelengths and polarization angles. These images demonstrate how the focal point of the metalens shifts to different z-positions as the input polarization is rotated. A comparison between the experimental and theoretical PSF, as determined from scalar diffraction theory, is shown in Fig.~\ref{fig:sim_compare}. 

To characterize each PSF, we fit each XY-plane within a 3D image stack to a two-dimensional un-normalized Gaussian function:
 \begin{equation}
 \label{eqn:gaussfit}
     I(x,y) = A\exp{\left( \frac{(x - x_0)^2 + (y - y_0)^2}{2\sigma^2} \right) .}
\end{equation}
The focal plane of the metalens associated with a given input wavelength and polarization was defined as the XY-plane yielding the greatest fitted Gaussian amplitude $A$. To estimate the depth of field of the metalens, we determined the axial depth of field as the full-width-at-half-maximum (FWHM) of fitted values of $A$ at different XY-planes. The lateral FWHM was determined from the Gaussian fit of the intensity distribution in the focal plane. 

The metalens focal length, depth of field, and lateral FWHM are plotted versus wavelength and polarization in Fig.~\ref{fig:2}(b). The metalens exhibits a tunable focal length ranging from approximately $f_{\rm min} = 220~{\rm \upmu m}$ ($\mathit{NA} = 0.09$) to $f_{\rm max}= 550~{\rm \upmu m}$ ($\mathit{NA} = 0.04$). This corresponds to a relative tuning range of $\left(f_{\rm max} - f_{\rm min}\right)/f_{\rm max}{\approx}55\%$. This tuning range is comparable to that realized with metasurfaces on elastically-deformable substrates~\cite{Zhan2017MetasurfaceNanophotonics}.  

A few features of our metalens differ from those of an ideal lens. The lateral FWHM values are close to the values predicted from an ideal diffraction-limited lens, ${\sim}\lambda / \left(2\mathit{NA}\right)$. However, the focal-plane intensity distributions, Fig.~\ref{fig:2}(c), feature slightly narrower central spots and more pronounced side rings than would be predicted by an ideal Airy disk. This discrepancy is attributed to a combination of greater optical transmission towards the edges of the metasurface aperture \cite{Rivolta1986AiryRatio}, phase aberrations resulting from imperfect fabrication of nanopost dimensions, and shifting of the focal spot since the ratio $\mathit{NA}\cdot D/\lambda$ is small (see Sec.~\ref{sec:SI1C} for additional details). In addition, the axial depth of field is asymmetric about the focal plane, and two to three times larger than would be predicted by the Rayleigh length, $z_R{\approx}\lambda / \left(\pi\mathit{NA}^2\right)$.  

The combined metalens and output polarizer system features a $180\degree$ polarization rotational symmetry with a sharp discontinuity in focal length about a particular wavelength-dependent axis. This discontinuity axis is typically $\theta{\approx}180\degree$ for wavelengths $\lambda{\approx}480\mbox{--}580~{\rm nm}$ and $\theta{\approx}150\mbox{--}180\degree$ for wavelengths $\lambda{\approx}580\mbox{--}650~{\rm nm}$. For polarization values near the discontinuity axis, two distinct local maxima in the PSF are observed due to Fresnel diffraction~\cite{JosephW.Goodman1996IntroductionOptics,Khorasaninejad2017AchromaticDispersion}. Since we define focal length as the location of maximum intensity, when the global maximum switches from the first to the second local maximum, an apparent discontinuity in the focal tuning appears. 

We also characterized the transmission and focusing efficiency properties of the optical system at the representative wavelength $\lambda = 581$ nm. Using the experimental setup shown in Fig.~\ref{fig:transmission}, we recorded three efficiency-related metrics of the optical system as a function of input light polarization: (i) the optical transmission through the metalens, (ii) the transmission through the metalens and output polarizer, and (iii) the combined system's focusing efficiency, defined as the ratio of optical intensity measured within the focal spot to the input optical intensity. These metrics are plotted in Fig.~\ref{fig:2}(d). The metasurface exhibits high transmission, $\gtrsim85\%$, independent of input polarization. However, addition of the output polarizer reduces the overall transmission and focusing efficiency in a polarization-dependent fashion. The focusing efficiency varies from $\sim50\%$, when the input optical polarization is aligned with the output polarizer, to $\lesssim5\%$ when the input polarization is orthogonal to the output polarizer axis. This variation may be a liability for photon-limited applications. Further work will investigate whether optimizing nanopost orientations~\cite{Arbabi2015DielectricTransmission,BalthasarMueller2017MetasurfacePolarization}, in addition to aspect ratios, may improve the focusing efficiency.  

\begin{figure*}[htbp]
    \centering
\includegraphics[width=.98\textwidth]{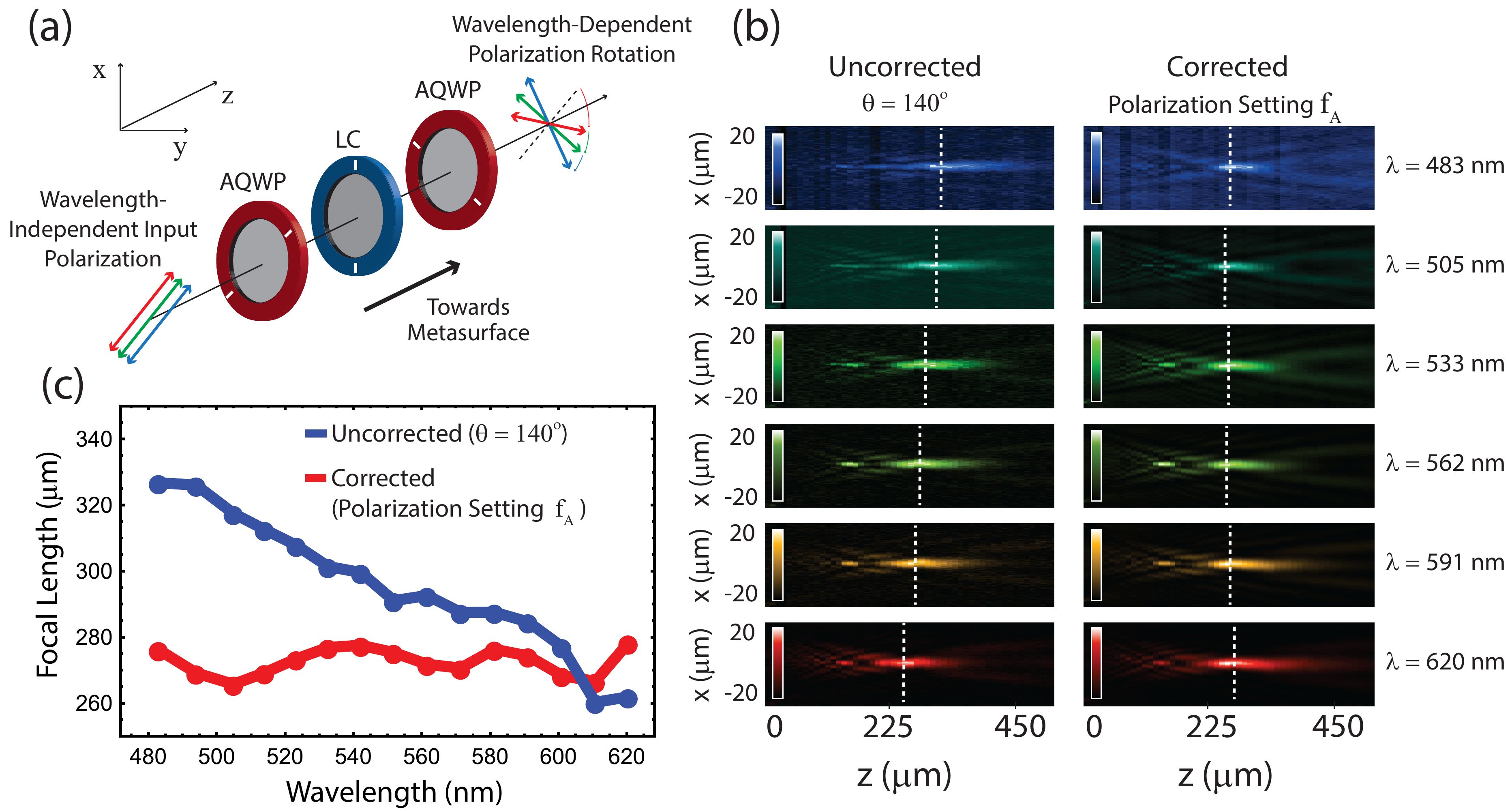}\hfill
       \caption{\label{fig:3}
\textbf{A polarization rotator corrects chromatic aberration.} a) A polarization rotator, consisting of a liquid-crystal variable retarder (LC) sandwiched between two achromatic quarter wave plates (AQWP), imparts a wavelength-dependent polarization rotation on linearly-polarized input light. b) XZ-slices for different wavelengths using two different input polarizations settings: (left) constant input polarization $\theta = 140\degree$ and (right) chromatically-corrected focal length setting $f_A$. Fitted focal lengths are denoted as vertical dashed lines. Intensity scale bars are normalized to largest value in each panel. c) Fitted focal length of the metalens for constant input polarization $\theta = 140\degree$ (uncorrected), and the $f_A$ setting (corrected). Applying the correction reduces the focal length variation by a factor of ${\sim}5$. 
       }
\end{figure*}

\section{{\label{subsec:achro}Correction of axial chromatic aberration}}
Many metalens designs exhibit prominent chromatic aberrations, similar to those encountered when using conventional diffractive optics. To find widespread application in imaging and display devices, achromatic metasurface optics are desirable. Recently, achromatic metasurfaces have been demonstrated that use dispersion engineering \cite{Arbabi2017ControllingMetasurfaces,Khorasaninejad2017AchromaticDispersion}, spatial multiplexing \cite{Arbabi2016MultiwavelengthMultiplexing,Lin2016PhotonicArray}, complex meta-atom geometries \cite{Chen2018AVisible,Shrestha2018BroadbandMetalenses,Chen2019ANanostructures}, and applications of the Pancharatnam-Berry phase \cite{Chen2018AVisible,Wang2018AVisible}. 

Inspection of Fig.~\ref{fig:2}(b) reveals an alternative route to realizing achromatic metalens behavior. In our metalens system, curves of constant focal length can be constructed by allowing the input optical polarization to vary as a function of wavelength. 

We implemented the requisite polarization rotations using the optical system shown in Fig.~\ref{fig:3}(a). The wavelength-dependent polarization rotator (WDPR) consists of an achromatic quarter waveplate oriented with fast axis at 45$\degree$, a liquid-crystal tunable retarder with slow axis at 0$\degree$, and a second achromatic quarter waveplate oriented at 135$\degree$. Together, these elements rotate the polarization of linearly-polarized incident light an amount $\Delta\theta(\lambda)$, which depends on the optical path length $\mathit{OPL}$ along the liquid-crystal's slow axis as:
\begin{equation}
\label{eqn:rotator}
    \Delta\theta(\lambda) = \frac{\pi \mathit{OPL}}{\lambda} .
\end{equation}
Since $\mathit{OPL}$ can be controlled by adjusting a calibrated voltage to the liquid crystal, this design allows for varifocal achromatic metalens behavior without macroscopic mechanically-moving parts. A derivation of Eq.~\eqref{eqn:rotator} is detailed in Sec.~\ref{sec:SIrotator}.

We selected two achromatic focal length settings $(f_A = 278~{\rm \upmu m}, f_B = 375~{\rm \upmu m})$ for further study. These focal lengths were set by adjusting the input polarization and the voltage of the liquid-crystal retarder. For both settings, the light output by the WDPR was characterized using a polarimeter. The results of these measurements are shown as an annotation in Fig.~\ref{fig:2}(b). We observe that the measured $\Delta\theta(\lambda)$ for both settings follows unique contours of constant focal length. 

To confirm that an achromatic correction has been realized, we acquired the metalens PSF at various wavelengths using the WDPR settings for $f_A$. Figure~\ref{fig:3}(b) shows the resulting XZ-slices of the PSF. As a reference, the XZ-slices for constant input polarization of $\theta = 140\degree$ are also shown. The corresponding fitted focal lengths are plotted as a function of wavelength in Fig.~\ref{fig:3}(c). For the constant $\theta=140\degree$ polarization setting, the focal length varies from $326\mbox{--}260~{\rm \upmu m}$ over the wavelength range $483\mbox{--}620~{\rm nm}$.  In comparison, using the achromatic focal length setting $f_A$, the focal length variation is five times smaller ($265\mbox{--}278~{\rm \upmu m}$) over the same wavelength range. 

The settings on the WDPR were not adjusted throughout these measurements, so the results in Figs.~\ref{fig:3}(b-c) show the potential of this approach for passive achromatic focusing of white light. In principle, for a given achromatic focal-length setting, a fixed waveplate with the appropriate wavelength-dependent retardance could be substituted for the liquid crystal. However, the voltage-tunable retarder enables rapid variation of the (chromatically corrected) focal length without any interchange of waveplates.

\begin{figure*}[htbp]
    \centering
\includegraphics[width=.98\textwidth]{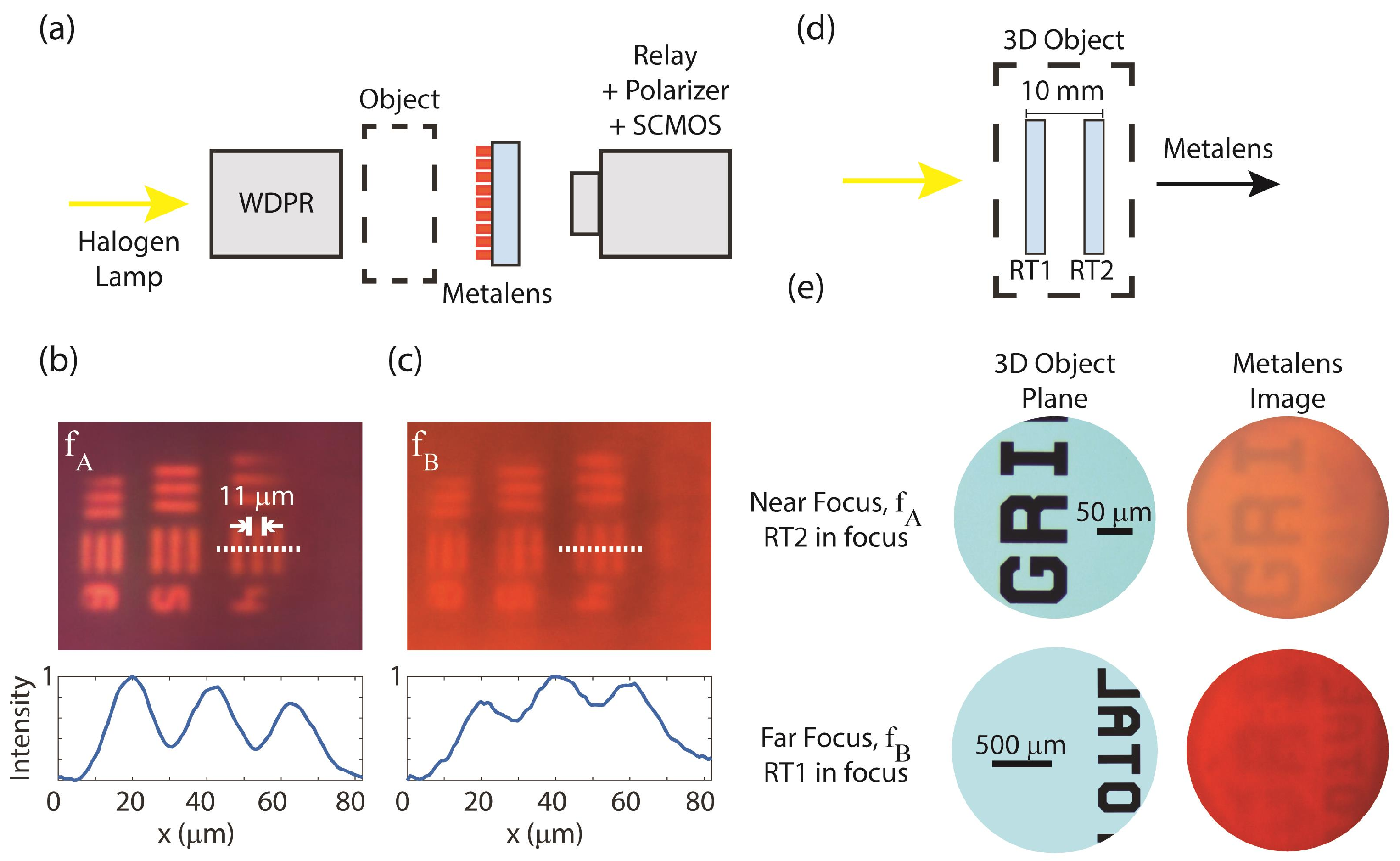}\hfill
       \caption{\label{fig:widefield}
\textbf{Broadband imaging demonstration.} a) Schematic of the experimental setup for widefield imaging. A halogen lamp source is routed through the wavelength-dependent polarization rotator (WDPR) to correct for chromatic aberration. The light subsequently illuminates an object, which is imaged by the metalens, magnified using a conventional-lens relay, and recorded with a color sensitive image sensor. b-c) (top) Widefield images of a negative (chrome background with transmissive elements) United States Air Force (USAF) resolution target imaged at two WDPR focal-length settings, $f_A$ and $f_B$. To maintain a constant image magnification, the distances between the resolution target, metalens, and relay were configured differently for the two focal length settings. (bottom) Linear intensity profile through the $11\mbox{-}{\rm \mu m}$-half-pitch line array in Group 5, Element 4. The resolution is higher for the $f_A$ setting due to its higher numerical aperture. d) Schematic of the 3D object used to demonstrate varifocal achromatic behavior. The object is comprised of two positive (transmissive background with chrome elements) resolution targets, labeled RT1 and RT2. The targets are separated by 10 mm using stacked glass slides coated in index-matched oil. The metalens is positioned such that RT2 is in focus using the $f_A$ setting and RT1 is in focus using the $f_B$ setting without any manual adjustment of the setup. Alternation between $f_A$ and $f_B$ settings is done electronically by changing the liquid-crystal voltage in the WDPR. e) (top row) Using the $f_A$ setting, only RT2 is in focus and the letters `GRI' are resolved. (bottom row) With the $f_B$ setting, the letters `OTAL' are visible on RT1, while the letters `GRI' on RT1 become blurry. Additional details on the imaging procedure are discussed in Sec.~\ref{sec:SIbroadband}. 
       }
\end{figure*}

\section{\label{subsec:whitelightimaging}Broadband imaging and focal switching}
Having demonstrated the ability to correct for chromatic aberration, we next used our varifocal metalens to perform widefield imaging with white light. The setup is illustrated in Fig.~\ref{fig:widefield}(a), and additional details are provided in Sec.~\ref{sec:SI3}. A resolution target was positioned ${\sim}1~{\rm mm}$ from the metalens aperture.
A halogen light source was shined through the WDPR [Fig.~\ref{fig:3}(a)] and used to back-illuminate the resolution target. The WDPR was controlled electronically to realize two achromatic focal length settings, $f_A$ and $f_B$. At each setting, an image of the resolution target was produced by the metalens. This image was magnified by 12x using a conventional-lens relay, and recorded using an RGB image sensor.  

Figure~\ref{fig:widefield}(b-c) shows acquired images of the resolution target at each of the two focal length settings. Lines separated by $11~{\rm \upmu m}$ are discernable for both $f_A$ and $f_B$ settings. However the resolution for the shorter focal length setting, $f_A$, is better than that of $f_B$ due to the higher effective numerical aperture of $f_A$. Images taken with the $f_B$ setting also exhibit higher background intensity and a different color than those taken with the $f_A$ setting. We attribute this background to undiffracted light from the halogen source reaching the camera sensor. At focal length $f_A$, this background light has a range of polarizations from $\theta=130\mbox{--}160\degree$, depending upon wavelength [Fig.~\ref{fig:2}(b)]. Hence, the output polarizer (oriented at 45$\degree$) removes the majority of the background illumination. For setting $f_B$, the background light has a polarization more closely aligned with the output polarizer, causing increased background transmission. 

To further demonstrate the metalens' achromatic varifocal behavior, we imaged a 3D object consisting of two positive resolution targets separated by 10 mm, Figure~\ref{fig:widefield}(d). Each resolution target was selectively imaged by alternating the WDPR settings between $f_A$ and $f_B$. For this experiment, the distances between the object, metalens, and optical relay were kept constant and variation between the two focal planes was realized entirely by electronic control of the WDPR.

Figure~\ref{fig:widefield}(e) shows the color images taken at each of the two WDPR settings. In the $f_A$ setting, the near target appears in focus with a de-magnification factor of 1.9. Under this setting, the far resolution target is blurry and barely visible as it lies far outside of the imaging system's depth of field. In the $f_B$ setting, the far resolution target appears in focus with a de-magnification factor of 13.2. In this setting, the near resolution target is blurry but still visible, owing to the larger depth of field. The image contrast for the $f_B$ setting was worse than for the $f_A$ setting, owing to transmitted undiffracted light.  For this image, we subtracted the background intensity [Sec.~\ref{sec:SIbroadband}] to partially improve the image contrast. All other images in Fig.~\ref{fig:widefield} correspond to the raw image-sensor RGB output. 

\section{\label{sec:summary}Outlook and conclusion}
Our results highlight the capability of birefringent metalenses to achieve achromatic, varifocal behavior by varying the polarization of incident light. Polarization-based focal tuning obviates the need to include electrodes, a deformable supportive mesh, or mechanical translators into the optical system. With increased numerical aperture, our approach may be ideally suited for realizing closely-packed arrays of varifocal metalenses for applications in computational microscopy and imaging \cite{Ng2005LightCamera,Levoy2006LightMicroscopy,Broxton2013WaveMicroscope}. Even with the existing NA (ranging from 0.04-0.09), our design may find application in novel light-field imaging architectures \cite{Lin2019AchromaticImaging}. Similar metalens designs may also be suitable for applications in computational displays \cite{Shi2018Polarization-dependentDisplays}. Existing compact displays use stacked layers of liquid crystal material to manipulate polarization. Hence, a polarization-based means of focal-length adjustment could be readily incorporated into existing display architectures. 

One drawback of our metalens design is its low focusing efficiency ($5\mbox{--}50\%$). This is a consequence of the use of an output polarizer. In future work, we will investigate more complex nanopost geometries \cite{Chen2018AVisible,Shrestha2018BroadbandMetalenses,Chen2019ANanostructures} that can be used to improve focusing efficiency without compromising focus tunability. Another drawback is that, while our fabrication technique allows for much larger metasurfaces ($\gtrsim1~{\rm mm^2}$, see Sec.~\ref{sec:sem}), we restricted the present study to a metalens with a relatively small aperture diameter ($D = 40~{\rm \upmu m}$). This is because, in our design, continuous focal tuning is only achieved for metalenses possessing $\lesssim1~{\rm radian}$ of phase difference between $H$ and $V$ polarized light, see Sec.~\ref{sec:SI1B}. This requirement places significant constraints on both the numerical aperture and diameter of the metalens. To circumvent this issue, a tunable metasurface could be combined with a conventional (refractive) lens \cite{Chen2018BroadbandOptics}. Alternatively, multiple tunable metasurfaces could be cascaded into a single device, with polarization rotators inserted between successive elements. Computationally-optimized freeform metasurface geometries \cite{Niederberger2014SensitivityAdjoints,Sell2017Large-AngleGeometries} may also improve the lens design. 

In summary, we designed and fabricated a metalens that has a polarization-tunable focal length and showed that applying a wavelength-dependent polarization rotation on incident light can correct for axial chromatic aberration. We characterized the 3D PSF of the metalens as a function of polarization and wavelength and used the metalens to perform broadband, widefield imaging with nearly diffraction-limited performance. Our results hold promise for the use of metalenses in future display and imaging applications.

\begin{acknowledgments}
We gratefully acknowledge advice and support from F. Hubert, Y. Silani, I. Brener, A. James, and J. Nogan and the use of nanofabrication and characterization resources at the Department of Energy Center for Integrated Nanotechnologies (CINT).

\textbf{Competing interests.} M. D. Aiello, A. S. Backer, and V. M. Acosta are co-inventors on a related pending patent application. J. D. Perreault and P. Llull are employed by Google LLC. The remaining authors declare no competing financial interests.

\textbf{Author contributions.} M. D. Aiello fabricated metasurfaces, developed optical characterization instruments, acquired and analyzed data. A. S. Backer designed the metasurfaces and polarization rotator, developed optical characterization instruments, acquired and analyzed data. A. J. Sapon performed control measurements and analyzed data. J. Smits developed instrumentation software and analyzed data. J. D. Perreault and P. Llull contributed to the experimental design and interpretation. V. M. Acosta supervised the project and assisted with data collection and analysis. All authors contributed to planning of experiments, discussion of results, and writing of the manuscript. 

\textbf{Funding.} This work was supported by a Google VR Faculty Research Award and a Google Daydream University Research Award. A. S. Backer acknowledges support from the Harry S. Truman Fellowship. This work was supported by the Laboratory Directed Research and Development program at Sandia National Laboratories, a multi-mission laboratory managed and operated by National Technology and Engineering Solutions of Sandia, LLC, a wholly owned subsidiary of Honeywell International, Inc., for the US Department of Energy's National Nuclear Security Administration under contract [DE-NA0003525]. This paper describes objective technical results and analysis. Any subjective views or opinions that might be expressed in the paper do not necessarily represent the views of the US Department of Energy or the United States Government. 

\end{acknowledgments}

\bibliographystyle{apsrev4-1}

\widetext
\clearpage

\begin{center}
\textbf{\large Supplemental Information}
\end{center}
\setcounter{equation}{0}
\setcounter{section}{0}
\setcounter{figure}{0}
\setcounter{table}{0}
\setcounter{page}{1}
\setcounter{equation}{0}
\setcounter{figure}{0}
\setcounter{table}{0}
\setcounter{page}{1}
\makeatletter
\renewcommand{\thetable}{S\arabic{table}}
\renewcommand{\theequation}{S\arabic{equation}}
\renewcommand{\thefigure}{S\arabic{figure}}
\renewcommand{\thesection}{S\Roman{section}}
\renewcommand{\bibnumfmt}[1]{[S#1]}
\renewcommand{\citenumfont}[1]{S#1}

%\begin{bibunit}

\section{Simulation and Design}

In this section, we provide the details of the design process that we used to realize a varifocal metalens. First, we overview the finite difference time domain simulations that were used to determine the TiO$_2$ nanopost dimensions that compose the metalens. Next, we discuss how fully continuous tunability is achieved by rotating the input polarization, as well as the design constraints that this varifocal feature necessitates. To verify our design and fabrication, we compare our experimentally measured PSF with the PSF predicted by Fourier optics simulations. Finally, we detail our implementation of a wavelength-dependent polarization rotator, used for correction of axial chromatic aberration.   

\subsection{Design of nanopost dimensions using finite difference time domain simulations}
\label{sec:SI1A}
In order for our metasuface to achieve a specific, polarization-dependent optical response, we must appropriately select the dimensions of the individual TiO$_2$ nanoposts that compose the basic elements of our design. Finite difference time domain simulations were carried out using the commercial software package FDTD Solutions (Lumerical). A single rectangular TiO$_2$ nanopost of 600 nm height was simulated on a quartz substrate. The refractive index of the TiO$_2$ material was incorporated using ellipsometry data reported in \cite{siDevlin2016BroadbandSpectrum.}. H- and V-polarized normally incident plane waves were injected through the substrate, and permitted to propagate along the z-axis. Periodic boundary conditions (400-nm pitch) were used along the x- and y-dimensions in order to model the effects of an array of nanoposts of similar dimensions. After each simulation, both the H- and V-polarized phase were measured in a plane 500 nm after the nanopost (using the monitor function provided in FDTD Solutions), and stored in a lookup-table. Simulations were repeated for nanopost lengths and widths ranging from 75-350 nm, varying each parameter in 25 nm increments. Hence, a total of 144 distinct simulation trials were performed in a two-dimensional parameter sweep. Phase values were re-offset relative to the phase recorded for a 175-by-175 nm square nanopost to produce the lookup tables plotted in Figs.~\ref{fig:idea}(c-d). For each lattice position in the metalens, these lookup tables were compared to the desired phase plotted in Fig.~\ref{fig:idea}(b) to select the most appropriate nanopost lengths and widths to match the desired H- and V-polarized phase response. Specification of the nanopost dimensions was carried out using a design wavelength of 581 nm. Hence, the phase response at other wavelengths does not necessarily match the ideal H- and V-polarized responses. This effect contributes to wavelength-dependent variation of the focal length and additional optical aberrations. However, as we demonstrate in Fig.~\ref{fig:3} of the main text, axial chromatic aberrations may be mitigated by rotating the input polarization in a wavelength-dependent manner.     

\subsection{Analysis of polarization-dependent metalens aperture phase}
\label{sec:SI1B}
In this section, we further discuss how our metalens achieves a continuously tunable focal length. The polarization-dependent phase response of the complete optical system can be modeled using Jones matrices. We denote purely H- and V-polarized phase responses at a spatial location $(x,y)$ as $e^{i\phi_H(x,y)}$ and $e^{i\phi_V(x,y)}$ respectively. The following calculation uses complex electric fields, though only the intensity (squared modulus) of the electric field is observed in experiments. 

The input electric field vector incident upon the metasurface is denoted as $\mathbf{E}_{in}^{T} = \left[\cos{\left( \theta \right)},\sin{\left( \theta \right)}\right]$. The output electric field vector, $\mathbf{E}_{out}$, is computed as:

 \begin{equation}
 \label{eqn:jonespol}
 \begin{split}
     \mathbf{E}_{out} & = \mathbf{J}_{2}\mathbf{J}_{1}\mathbf{E}_{in} \\
     & = \frac{1}{2} \begin{bmatrix}
     1 & 1  \\
     1 & 1  
     \end{bmatrix}
     \begin{bmatrix}
     e^{i\phi_H(x,y)} & 0  \\
     0 & e^{i\phi_V(x,y)}  
     \end{bmatrix}
     \begin{bmatrix}
     \cos{\left(\theta\right)}  \\
     \sin{\left(\theta\right)}  
     \end{bmatrix} \\
     & = \frac{1}{2} \left(\cos{\left(\theta\right)}e^{i\phi_H(x,y)}+\sin{\left(\theta\right)}e^{i\phi_V(x,y)}\right)\begin{bmatrix}
     1 \\
     1
     \end{bmatrix} .
 \end{split}
 \end{equation}

In Eq.~\eqref{eqn:jonespol}, $\mathbf{J}_{1}$ and $\mathbf{J}_{2}$ denote Jones matrices representing the metasurface and output polarizer respectively. We conclude that the output electric field will be polarized at 45$\degree$, while the output phase is determined as: $\phi_{out}(x,y) = \angle \left(\cos{\left(\theta\right)}e^{i\phi_H(x,y)}+\sin{\left(\theta\right)}e^{i\phi_V(x,y)}\right)$, where the symbol $\angle$ denotes the phase of the expression in parentheses. We note that a rotation of the input polarization $\theta$ induces a change in the phase  $\phi_{out}(x,y)$ output by the metalens/output polarizer combination. In contrast, if an output polarizer were not included in our experimental system, the output intensity would simply be a linear combination of H- and V-polarized input intensities, but the output phases of the individual H- and V-polarized electric field components would not be affected. From Eq.~\eqref{eqn:jonespol}, the scalar output electric field $E_{out}(x,y)$ immediately after the metalens aperture plane is expressed as:

\begin{equation}
\label{eqn:jonespol_scalar}
    E_{out}(x,y) = \frac{1}{2} \left(\cos{\left(\theta\right)}e^{i\phi_H(x,y)}+\sin{\left(\theta\right)}e^{i\phi_V(x,y)}\right) .
\end{equation}

For brevity, we no longer explicitly represent the $45\degree$ polarization of the output field. We derive an approximate expression for the phase immediately after the metalens aperture as a function of the input polarization $\theta$. Throughout this analysis, we assume that $\phi_H(x,y)$ and $\phi_V(x,y)$ are parabolic functions associated with lenses of focal lengths $f_H$ and $f_V$ respectively:

\begin{equation}
\label{eqn:focal_phase}
    \phi_{H,V}(x,y) = \frac{-\pi}{\lambda f_{H,V}}\left(x^2 + y^2\right)
\end{equation}

By denoting the difference in phase between $\phi_H(x,y)$ and $\phi_V(x,y)$ as $2\delta(x,y)$, we re-write the electric field distribution just after the metasurface aperture as:

\begin{equation}
\label{eqn:del_psi}
     E_{out}(x,y) = \frac{1}{2} e^{i\psi(x,y)} \left(\cos{\left(\theta\right)}e^{i\delta(x,y)}+\sin{\left(\theta\right)}e^{-i\delta(x,y)}\right) ,
\end{equation}
where:

\begin{equation*}
    \psi(x,y) =  \frac{-\pi\left(f_H + f_V\right)}{2\lambda f_H f_V}\left(x^2 + y^2\right) ,
\end{equation*}
and:

\begin{equation*}
    \delta(x,y) =  \frac{-\pi\left(f_V - f_H\right)}{2\lambda f_H f_V}\left(x^2 + y^2\right) .
\end{equation*}

To simplify the expression for $E_{out}(x,y)$ further, we now make an approximation. Assuming that the phase factor $\delta(x,y)$ is small ($\ll \pi$), we use the first-order Taylor expansion:

\begin{equation}
\label{aprx:delta}
    e^{i\delta(x,y)} \approx 1 + i\delta(x,y) .
\end{equation}

Substituting Approximation~\eqref{aprx:delta} into Eq.~\eqref{eqn:del_psi} and simplifying, we obtain:

\begin{equation*}
     E_{out}(x,y) = \frac{1}{2} \left(\cos{\left(\theta\right)} + \sin{\left(\theta\right)}\right) e^{i\psi(x,y)} \left(1 + i\gamma(\theta)\delta(x,y)\right) ,
\end{equation*}

where:

\begin{equation}
\label{eqn:alpha}
   \gamma(\theta) = \frac{\left(\cos{\left(\theta\right)} - \sin{\left(\theta\right)}\right)}{\left(\cos{\left(\theta\right)} + \sin{\left(\theta\right)}\right)} .
\end{equation}

In order to to represent the phase response of the metasurface in an easy to analyze form, we now consider the limit when the product $\gamma(\theta)\delta(x,y)\ll \pi$. This simplification allows us to apply another Taylor approximation similar to that of Approximation \ref{aprx:delta}. As discussed later in this section, this limit is not satisfied for all values of $\theta$. We use it here only to gain intuition about the metasurface phase response, but all simulations performed in the main text use the more rigorous approach described in Sec.~\ref{sec:SI1C}. Nevertheless, in this limit, $E_{out}(x,y)$ simplifies to:

\begin{equation}
\label{eqn:Eout_aprx}
    E_{out}(x,y) = \frac{1}{2} \left(\cos{\left(\theta\right)} + \sin{\left(\theta\right)}\right) e^{i\left(\psi(x,y) + \gamma(\theta)\delta(x,y)\right)} .
\end{equation}

Hence, the phase at the metasurface aperture is:

\begin{equation}
\label{eqn:phase_ap_aprx}
    \psi(x,y) + \gamma(\theta)\delta(x,y) = \frac{-\pi}{\lambda}\left(x^2 + y^2\right) \frac{\left(1 - \gamma(\theta) \right)f_H + \left(1 + \gamma(\theta) \right)f_V}{2 f_H f_V} .
\end{equation}

Equation~\eqref{eqn:phase_ap_aprx} shows that, assuming that the first-order Taylor expansions discussed above are valid, rotating the input polarization $\theta$ changes the value of $\gamma$ which, in turn, modulates the effective focal length of the metalens. By comparison with Eq.~\eqref{eqn:focal_phase}, it can be seen that Eq.~\eqref{eqn:phase_ap_aprx} is the phase factor of a lens of focal length $f_\theta$, given by:

\begin{equation}
\label{eqn:f_theta}
    f_\theta =  \frac{2 f_H f_V}{\left(1 - \gamma(\theta) \right)f_H + \left(1 + \gamma(\theta) \right)f_V} .
\end{equation}

In order to maintain the validity of our approximations, $\delta(x,y)$ must remain small across the entire metalens aperture. Heuristically, we have found that $\delta(x,y)$ should not exceed approximately one radian. This stipulation requires our metalens designs to have small aperture diameters ($D\approx40~{\rm \mu m}$) and relatively low numerical aperture ($NA\lesssim0.1$). Although this constraint limits the maximum-achievable focal tuning range, it does not prevent one from combining this metalens design with additional polarization-independent optics. For example, a metalens doublet containing a polarization-tunable surface and a polarization-independent surface could be used to achieve high numerical aperture focusing in addition to continuous focal tunability.

\begin{figure}[htbp]
    \centering
\includegraphics[width=.98\textwidth]{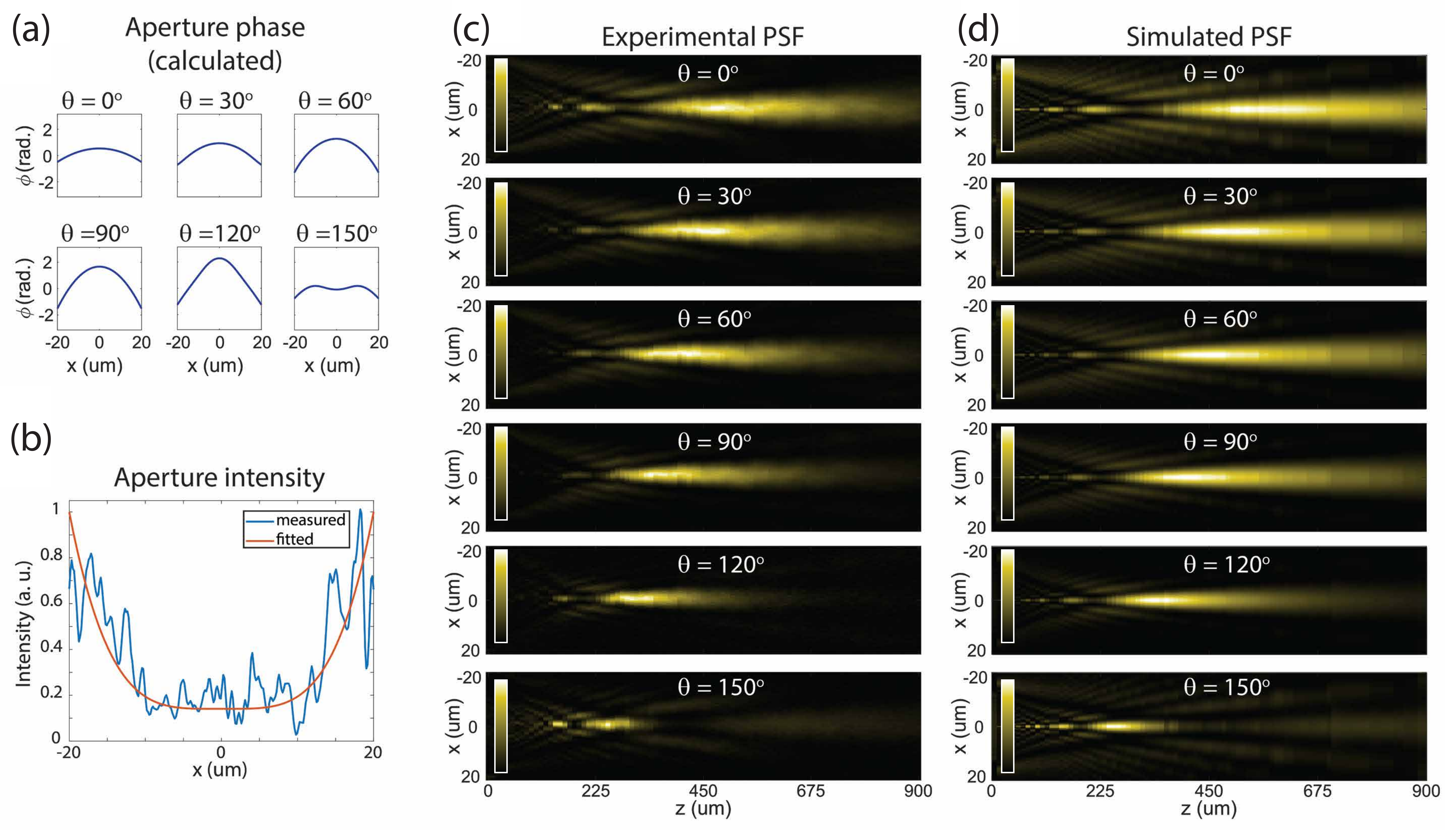}\hfill
       \caption{\label{fig:sim_compare}
\textbf{Simulation of the metalens PSF.} a) The calculated metalens aperture phase for different input polarizations $\theta$. b) Measured aperture plane intensity, and a fitted quartic function. c) XZ-slices of the measured PSF at the design wavelength $\lambda = 581$ nm. d) XZ-slices of the simulated PSF. Intensity scale bars are normalized to the largest value in each panel.  
       }
\end{figure}

We further point out that in order for Eq.~\eqref{eqn:Eout_aprx} and Eq.~\eqref{eqn:phase_ap_aprx} to remain valid, the product $\gamma(\theta)\delta(x,y)$ must also be small across the metalens aperture. However, for the input polarization $\theta = 135\degree$, the magnitude of $\gamma(\theta)$ will be infinite [see Eq.~\eqref{eqn:alpha}]. Accordingly, at input polarizations close to $\theta = 135\degree$, the aperture phase does not resemble the parabolic shape of a lens. To examine this effect, Fig.~\ref{fig:sim_compare}(a) computes the metalens aperture phase, without approximation, at the design wavelength $\lambda = 581$ nm, by numerically evaluating Eq.~\eqref{eqn:jonespol_scalar}. For polarizations $0\degree\lesssim \theta \lesssim 90\degree$, the magnitude of $\gamma(\theta)$ is less than 1. Accordingly, the aperture phase is well approximated as a parabola of varying concavity. However, for polarizations $90\degree \lesssim \theta \lesssim 180\degree$, the aperture phase contains ripples that do not resemble a parabolic function. A consequence of this effect is that the resulting metalens PSF contains aberrations in the form of multiple axial focal spots [see Fig.~\ref{fig:sim_compare}(c)]. Due to the aberrations present in the PSF, the estimated focal length of the metalens, defined as the plane containing the greatest fitted amplitude $A$ [Eq.~\eqref{eqn:gaussfit}], can change abruptly as a function of input polarization [see for example, the upper right corner of the array plots in Fig.~\ref{fig:2}(b)]. Due to these effects, for our white-light imaging experiments, we chose our achromatic focal length settings to lie on contours [Fig.~\ref{fig:2}(b)] that avoided sudden changes in focal length. 

To conclude this section, we briefly examine how a polarization-sensitive metalens would function \emph{without} the use of an output polarizer oriented at 45$\degree$. If no output polarizer were included, the intensity at a given point $I(x,y,z)$ at some point $z$ beyond the metasurface aperture is:

\begin{equation}
    I(x,y,z) = \cos^2{(\theta)}|E_H(x,y,z)|^2 + \sin^2{(\theta)}|E_V(x,y,z)|^2 .
\end{equation}
In other words, $I(x,y,z)$ is a linear combination of the output intensities associated with purely H- and V-polarized fields $E_H(x,y,z)$ and $E_V(x,y,z)$. However, we note that the phase of the individual polarization components is unaffected. In contrast, our use of an output polarizer projects $E_H(x,y,z)$ and $E_V(x,y,z)$ onto a single axis, causing the (previously orthogonal) electric field components to interfere. This gives rise to a unique overall output phase, which we leverage to tune the effective focal length of our optical system. 

\subsection{Simulation of the polarization-dependent metalens PSF}
\label{sec:SI1C}

To further confirm the varifocal behavior of the fabricated metalens, we performed Fourier optics \cite{siJosephW.Goodman1996IntroductionOptics} simulations to calculate the expected 3D PSF as a function of input polarization. Given an expected electric field distribution at the output of the metalens aperture, $E(x,y,0)$, the electric field at any axial distance $z$ beyond the aperture plane is computed using the Fresnel propagator $\mathcal{F}\mathcal{R}$ as: 

\begin{equation}
\label{eqn:fresnel}
    E(x,y,z) = \mathcal{F}\mathcal{R}\{E(x,y,0), z\} = \frac{e^{i\frac{2\pi z}{\lambda}}}{i\lambda z}e^{\frac{i\pi \left(x^2 + y^2\right)}{\lambda z}}\int\int E(x',y',0) e^{\frac{i\pi \left(x'^2 + y'^2\right)}{\lambda z}}e^{\frac{-i2\pi\left(xx'+yy'\right)}{\lambda z}}dx'dy' .
\end{equation}
To determine the aperture-plane electric field to be used as input to the Fresnel propagator, an input polarization $\theta$ was selected, and the complex electric field was determined using Eq.~\eqref{eqn:jonespol_scalar}. The computed radial phases across the metalens aperture for different input polarizations are shown in Fig.~\ref{fig:sim_compare}(a) for the design wavelength $\lambda = 581$ nm. Note that the phase curvature changes as a function of polarization. The aperture-phase calculations shown in Fig.~\ref{fig:sim_compare}(a) do not rely on the approximations discussed in the previous section. Additionally, we found that the metalens exhibits non-uniform transmission across the aperture. The highest transmission efficiency was observed at the outer edges of the metalens. To incorporate this effect into our Fourier optics simulations, the normalized radial intensity profile across the metalens aperture was measured for the $V$ input polarization and fit to a quartic polynomial. The measured and fitted aperture intensities are plotted in Fig.~\ref{fig:sim_compare}(b). The input electric field $E(x,y,0)$ was scaled by the square-root of the fitted aperture intensity, before using Eq.~\eqref{eqn:fresnel} to generate simulated PSFs. Simulations were performed using custom software written in MATLAB (Mathworks). A comparison between the experimental and simulated PSF for the design wavelength $\lambda = 581$ nm is shown in Figs.~\ref{fig:sim_compare}(c-d). Our calculations demonstrate good agreement with the measured intensity distributions. Interestingly, the measured intensity distributions in the focal plane exhibit narrower PSFs with more pronounced side-lobes than would be expected from back-of-the-envelope Gaussian beam calculations \cite{siAnthonyE.Siegman1986Lasers}. This effect is due primarily to the non-uniform transmission across the metalens aperture. Specifically, since the outer annulus of the metalens exhibits the greatest transmission, the resulting PSF is weighted towards the higher spatial frequencies \cite{siRivolta1986AiryRatio}. 

An additional consequence of using an output polarizer in our optical design is the non-uniform dependence of transmission and focusing efficiency on input polarization, as seen in Fig.~\ref{fig:2}(d). Including the measured aperture transmission in Eq.~\eqref{eqn:fresnel}, we calculated the expected focusing efficiency at the representative wavelength $\lambda = 581$ nm and compared with experimental measurements, Fig.~\ref{fig:foc_compare}. As expected, the overall focusing efficiency is maximized when the input polarization is aligned with the output polarizer at $\theta=45\degree$.  

\begin{figure}[htbp]
    \centering
\includegraphics[width=.5\textwidth]{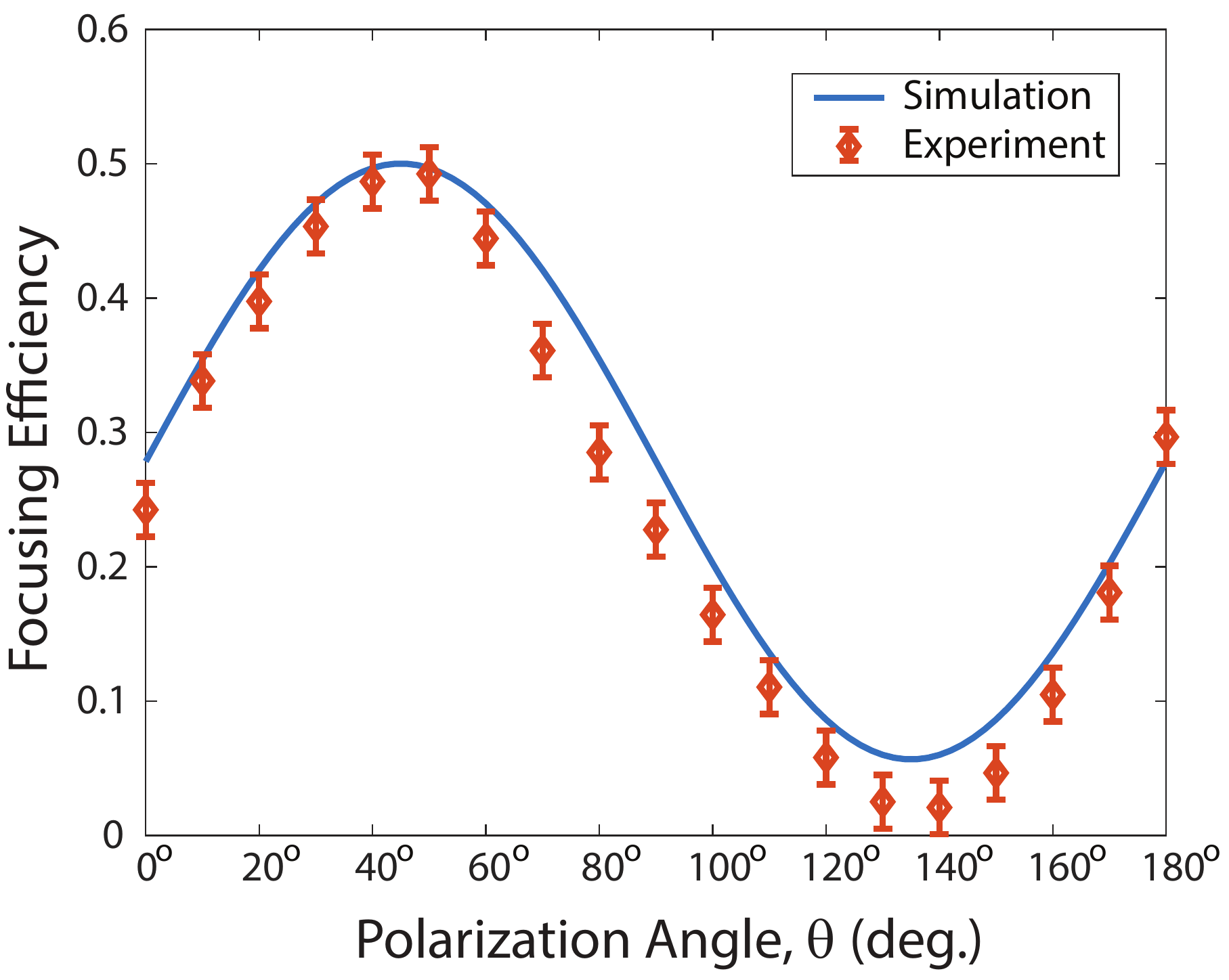}\hfill
       \caption{\label{fig:foc_compare}
\textbf{Comparison of experimental and simulated focusing efficiency.} Both simulations and experiments show a strong polarization-dependence. Maximum focusing efficiency is achieved when input and output polarizers are aligned at $\theta = 45\degree$, and minimum efficiency occurs when the input is orthogonal to the output polarizer at $\theta = 135\degree$. Since the experimental efficiency depends upon the transmission through the metasurface, in addition to other optical elements (e. g. polarizers, lenses etc.) within the experimental system, simulated focusing efficiency values were normalized to the maximum measured value ~\cite{siChen2018AVisible}. Error bars for the measured values show a measurement uncertainty of $\pm 0.02$, as derived from taking multiple efficiency measurements under the same settings and computing the standard deviations.   
       }
\end{figure}

\begin{figure}[htbp]
    \centering
\includegraphics[width=.5\textwidth]{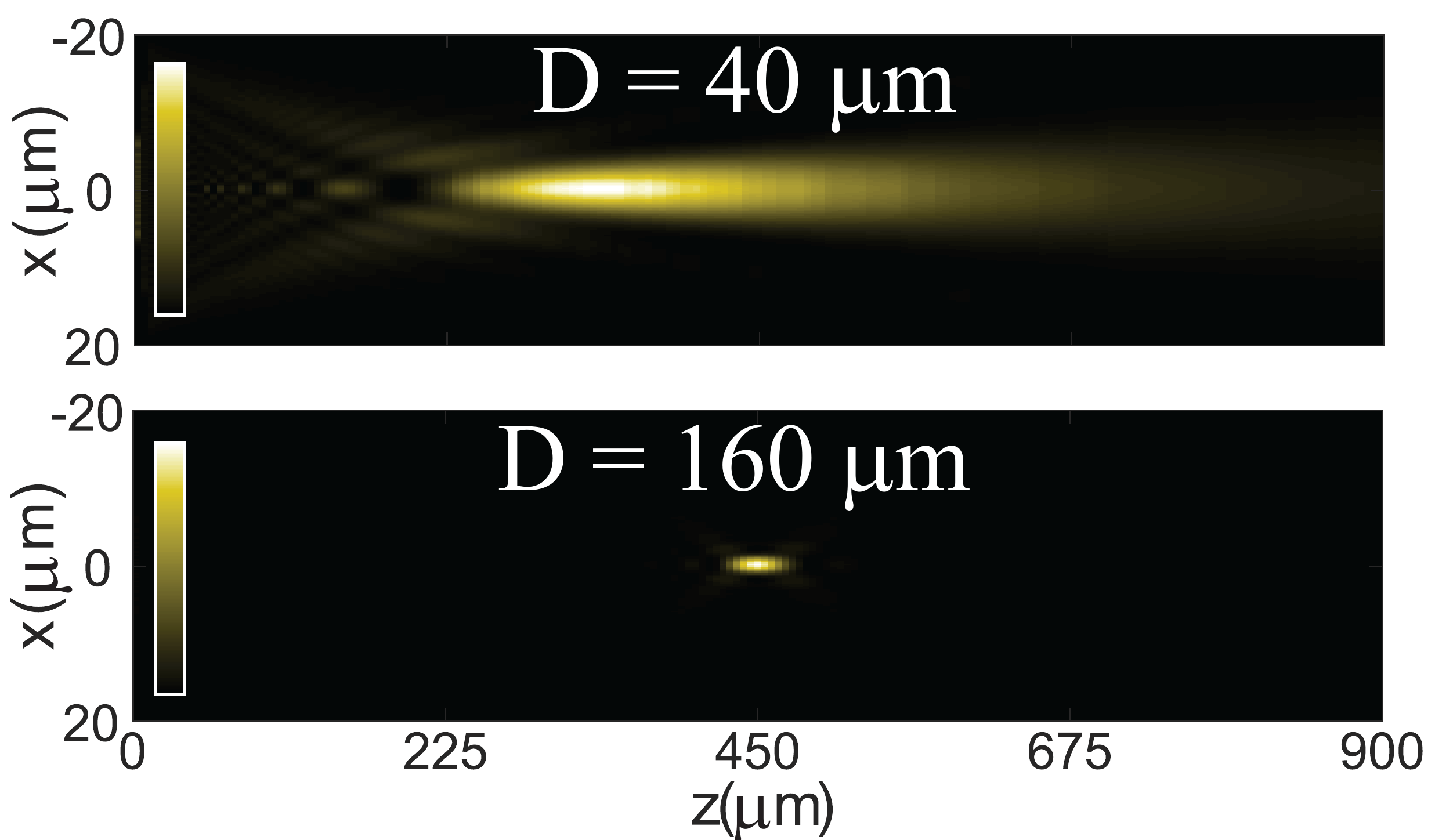}\hfill
       \caption{\label{fig:foc_compare}
\textbf{Simulation of lenses with different numerical apertures.} XZ-slices through the 3D PSF of two lenses are calculated using Eq.~\eqref{eqn:alpha_f}. Both lenses have the same design focal length, $f = 450~{\rm \upmu m}$, but different aperture diameters, $D = 40~{\rm \upmu m}$ ($\mathit{NA}=0.044$) and $D=160~{\rm \upmu m}$ ($\mathit{NA}=0.177$), respectively. The larger-diameter lens produces a focal spot near the designed focal length, whereas the smaller-diameter lens produces a focal spot significantly closer to the aperture.   
       }
\end{figure}

\subsection{Aperture size affects focal length}
\label{sec:SI1D}
In Sec.~\ref{sec:sim}, we noted that, within the paraxial approximation, the formula for the phase imparted by a thin lens of focal length $f$ is:

\begin{equation}
\label{eqn:thinlensSI}
    \phi(x,y) \approx -\alpha(f, \lambda)\cdot\left(x^2 + y^2\right) .    
\end{equation}
In the limit $\mathit{NA}\cdot D/\lambda>>1$, The relationship between $\alpha$ and the lens focal length $f$ is simply:

\begin{equation}
\label{eqn:alpha_f}
    \alpha(x,y) = \frac{\pi}{\lambda f} .    
\end{equation}
However, as the diameter of the lens aperture shrinks, Eq.~\eqref{eqn:alpha_f} is no longer a good predictor of the distance at which the axial focused intensity will reach a maximum. To illustrate this point, we used Eq.~\eqref{eqn:fresnel} to simulate two lenses with identical design focal length, $f=450~{\rm \upmu m}$, but different aperture diameters, $D = 40~{\rm \upmu m}$ and $D=160~{\rm \upmu m}$, respectively. XZ-slices of the simulated PSF of both lenses are shown in Fig.~\ref{fig:foc_compare}. Even though both lenses have the same phase variation across their apertures, the lens with smaller diameter appears to have a shorter focal length.

\subsection{Wavelength-dependent polarization rotator: Principle of operation}
\label{sec:SIrotator}
 In this section, we describe how the wavelength-dependent polarization rotator (WDPR) is implemented using two achromatic quarter waveplates (AQWP:  Thorlabs AQWPO5M) and an additional wavelength-dependent waveplate. In our experiments, a liquid-crystal retarder (LCR: Thorlabs LCC2415-VIS) was used as the wavelength-dependent waveplate, as this facilitated rapid switching between different (achromatic) focal-length settings.

The behavior of the WDPR is understood by analyzing the Jones matrices associated with the optical system. The effective Jones matrix of the two achromatic quarter waveplates and the wavelength-dependent waveplate, ($\textbf{J}_{rotator}$) is given by:

\begin{align}
    \nonumber
    \textbf{J}_{rotator} &= \textbf{J}_{135\degree} \textbf{J}_{LC} \textbf{J}_{45\degree}
\end{align}

where $\textbf{J}_{45\degree}$ and $\textbf{J}_{135\degree}$ are the Jones matrices associated with the AQWPs with fast-axis orientations aligned at $45\degree$ and $135\degree$ respectively. $\textbf{J}_{LC}$ is the Jones matrix associated with the liquid-crystal retarder acting as a wavelength-dependent waveplate aligned with fast-axis at $0\degree$. In this configuration, $\textbf{J}_{rotator}$ can be expressed as: 

\begin{align}    
    \nonumber
    \textbf{J}_{rotator} &= \frac{1}{4}
    \begin{bmatrix}
        1+i & -1+i  \\
        -1+i  & 1+i
    \end{bmatrix} 
    \begin{bmatrix}
        1       & 0  \\
        0       & e^{i2\Delta\theta(\lambda)} 
    \end{bmatrix}
    \begin{bmatrix}
        1+i  & 1-i  \\
       1-i  & 1+i  
    \end{bmatrix}\\
    \nonumber
     &= 
     ie^{i\Delta\theta(\lambda)} \begin{bmatrix}
     \cos(\Delta\theta(\lambda))       & -\sin(\Delta\theta(\lambda))  \\
     \sin(\Delta\theta(\lambda))       & \cos(\Delta\theta(\lambda)) 
     \end{bmatrix} \\
     \label{eqn:matrixrotator}
     &= ie^{i\Delta\theta(\lambda)}\textbf{R}(\Delta\theta(\lambda)) .
\end{align}

Here, the liquid-crystal retarder imparts a phase lag $2\Delta\theta(\lambda) = 2\pi\mathit{OPL}/\lambda$, where $\mathit{OPL}$ is the effective optical path length.  The resulting $\textbf{J}_{rotator}$ matrix, Eq.~\eqref{eqn:matrixrotator}, includes an overall phase shift $ie^{\Delta\theta(\lambda)}$ and, more importantly, a simple polarization rotation matrix, $\textbf{R}(\Delta\theta(\lambda))$, that depends on wavelength. 
 
 The overall function of this device is summarized as follows: Light incident upon the WDPR, $\mathbf{E}_{in}$, is linearly polarized with a polarization angle, $\theta$, that is independent of wavelength. The output after the WDPR, $\mathbf{E}_{out}$, is also linearly polarized, but different wavelengths are rotated different amounts. Hence:

\begin{align}
    \nonumber \mathbf{E}_{out} &= \textbf{J}_{rotator} \mathbf{E}_{in}\\
    \nonumber &= ie^{i\Delta\theta(\lambda)}
    \begin{bmatrix}
     \cos(\Delta\theta(\lambda))       & -\sin(\Delta\theta(\lambda))  \\
     \sin(\Delta\theta(\lambda))       & \cos(\Delta\theta(\lambda)) 
     \end{bmatrix}
     \begin{bmatrix}
            \cos(\theta)\\
            \sin(\theta)
     \end{bmatrix} \\
     \nonumber &= ie^{i\Delta\theta(\lambda)}
     \begin{bmatrix}
            \cos(\theta + \Delta\theta(\lambda))\\
            \sin(\theta + \Delta\theta(\lambda))
     \end{bmatrix}.
\end{align}

This shows that the output light is linearly polarized, but the polarization is rotated by the amount $\Delta\theta(\lambda)=\pi\mathit{OPL}/\lambda$.

Tuning the WDPR relies on manipulating two variables in the above calculation, $\theta$ and $\Delta\theta(\lambda)$. By setting $\theta$ we choose the wavelength-independent input polarization. We do this by using a linear polarizer mounted on a motorized rotation mount. To manipulate $\Delta\theta(\lambda)$ we adjust the voltage applied on the liquid-crystal retarder, and the effective $\mathit{OPL}$. This procedure permits us to control how much each wavelength is rotated in polarization space.

\section{Fabrication}
\label{sec:SIfab}

\begin{figure}[htbp]
    \centering
\includegraphics[width=.98\textwidth]{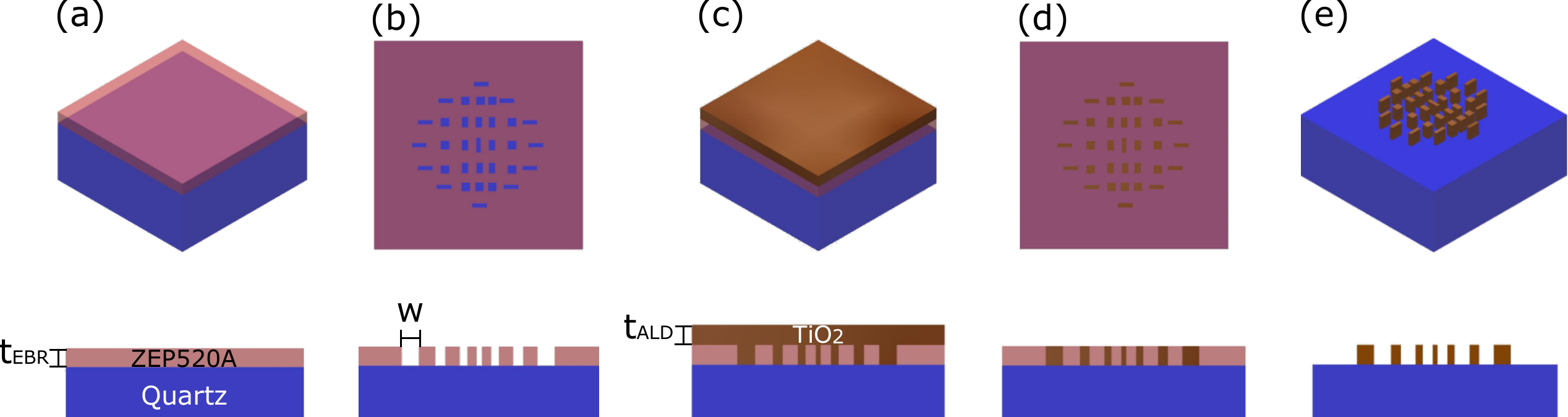}\hfill
       \caption{\label{fig:fab}
\textbf{Fabrication Process.} a) Spin coat the Electron Beam Resist (EBR) to thickness $t_{EBR}=600~{\rm nm}$. This thickness will set the height of the final nanostructures. b) Electron Beam Lithography (EBL) and development. This creates vias in the EBR which serves as a mask for the metasurface structures. $w$ represents the width of the widest vias in the pattern. c) Atomic Layer Deposition (ALD) coats the substrate with titanium-dioxide (TiO$_2$). ALD is used to conform to the side walls of the EBR, adhere to the quartz at the bottom of the vias, and to planarize the film once the vias are filled. A minimum deposition thickness of $w/2$ is required (i.e. $t_{ALD} \geq w/2$). d) Reactive Ion Etch (RIE) removes the excess TiO$_2$. Etching is stopped once the top of the EBR layer is reached. e) Final EBR removal leaves the nanostructures alone on the substrate and results in the final metasurface. 
       }
\end{figure}

Our metalens fabrication process is based on recently published methods  \cite{siDevlin2016BroadbandSpectrum.}. Our step-by-step approach is outlined in Fig.~\ref{fig:fab}. All fabrication was performed at Sandia National Labs Center for Integrated Nanotechnologies (CINT). First, we spincoat a 600 nm layer of ZEP520A positive electron beam resist (EBR) on top of a 1-mm-thick quartz substrate (Ted Pella, Inc. 26013). The EBR thickness sets the maximum achievable height of the nanoposts. Next, we perform a 20 nm thick thermal deposition of gold to serve as a dissipation layer for the electron beam lithography (EBL). We then expose the metasurface with our desired nanopost pattern using a JEOL EBL system operating at 1 nA beam current and 10 nm beam pitch. The thermal gold is removed after exposure using a potassium:potassium iodide:deionized water (4g:1g:50ml) mixture for 75 sec. Next, we perform atomic layer deposition (ALD) at $90\degree~{\rm C}$ to deposit an amorphous titanium dioxide (TiO$_2$) layer that fills the vias created by EBL. Using titanium tetrachloride (TiCl$_4$) and H$_2$O as precursors, we run 4000 cycles with pulse/purge times of $0.4~{\rm s}/6~{\rm s}$ for TiCl$_4$ and $0.4~{\rm s}/6~{\rm s}$ for H$_2$O. This process deposits a 350-nm-thick layer of TiO2. The thickness of this layer is chosen so that it is more than half the largest width dimension in the vias, to ensure complete filling. We then etch the undesired layer of TiO$_2$ on top of the EBR pattern until the tops of the EBR posts are reached. This is done using a Inductively Coupled Plasma Reactive Ion Etch (ICP-RIE) system. The specifics of the recipe include 12 standard cubic centimeters per minute (sccm) of boron trichloride (BCl$_3$), 20 sccm of chlorine (Cl$_2$), and 8 sccm of argon (Ar) at 350W RIE power and 35W ICP power. We have measured the etch rate of this process to be $0.85~{\rm nm/s}$. We monitor the etch using laser endpoint detection and stop the etch 10 seconds after we detect the change of etching TiO$_2$ to etching EBR. Finally, we soak the device in Remover PG at 70$\degree$C for 3 hours to dissolve the remaining EBR.

\begin{figure}[H]
    \centering
\includegraphics[width=.99\textwidth]{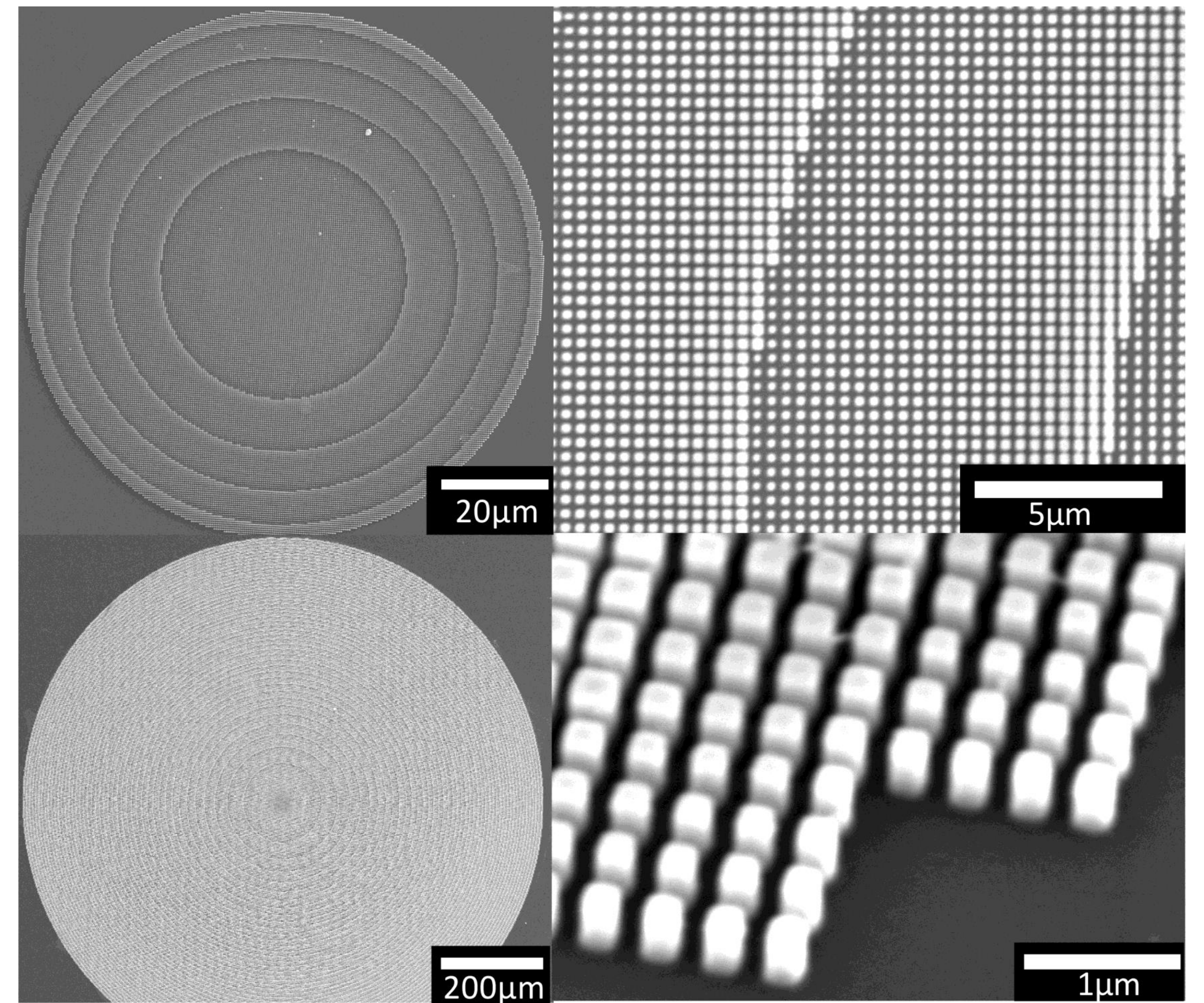}\hfill
       \caption{\label{fig:sem}
\textbf{Large diameter metalenses.} SEM images of larger-diameter metalenses. Top images show an SEM of a 100 $\mu$m metalens, bottom images show a 1 mm diameter metalens. 
        }       
\end{figure}

\section{\label{sec:sem}Additional Devices}

In this section we outline and describe addition metalens devices that we have fabricated using the process described in Sec.~\ref{sec:SIfab}. Metalenses with diameters as large as 1 mm and numerical aperture exceeding 0.4 were fabricated. Figure~\ref{fig:sem} shows SEM images of these larger diameter devices. The 1-mm-diameter metalenses required ${\sim}30~{\rm minutes}$ of EBL due to the need for writing more vias in the EBR. While scaling to moderately larger diameters may be realized by using higher EBL current, wafer-scale processes based on our metalens designs may require a mask-based deep UV photolithography process. Also, due to the larger numerical aperture and diameter of these devices, the design includes 'phase wraps.' This is where the phase output of the device transitions from $2\pi$ back to $0$. These phase wraps increase the effect of chromatic aberration due to each color requiring a 'wrap' at a different radius. However, for a single wavelength, the devices feature nearly diffraction-limited focusing behavior.

\section{\label{sec:Setup}Optical characterization and imaging}
\label{sec:SI3}
In this section, we provide details for the two experimental setups that were used to optically characterize the varifocal metalens. The first setup was used to measure the 3D point spread function (PSF) of the metalens as a function of polarization and wavelength. The second setup was used to perform varifocal color imaging of planar and 3D objects.

\begin{figure}[htbp]
    \centering
\includegraphics[width=.99\textwidth]{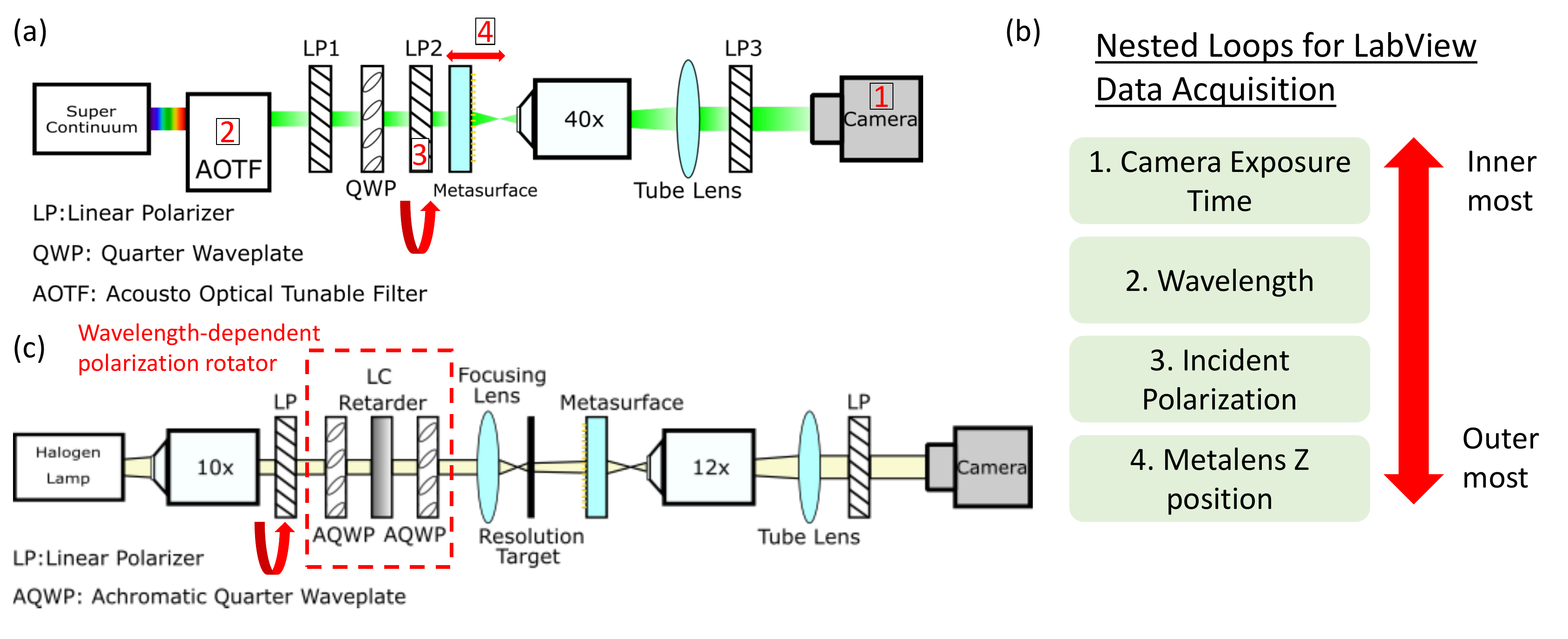}\hfill
       \caption{\label{fig:exp}
\textbf{Experimental Design and Automation.} a) PSF acquisition experimental setup. A supercontinuum source supplies spatially coherent light from 400-700+ nm. The light is passed through an Acousto Optical Tunable Filter (AOTF) to select a specific wavelength (5 nm bandwidth) incident on the metasurface. Subsequent polarization optics are used to choose the linear polarization incident on the metasuface by rotating LP2 with a motorized rotation mount. The 3D PSF is acquired by scanning the metasurface axially with a motorized translation stage. The light passing through the metasurface is imaged using a conventional-lens relay, passed through an output polarizer, and detected with a monochromatic image sensor. b) Flowchart outlining the LabView program used to take PSF data. The exposure time of the camera is adjusted to compensate for the wavelength-dependent intensity variation of the illumination light. The nesting order was chosen to optimize the speed of the measurement, while minimizing any instabilities associated with the metasurface axial scanning. c) Broadband imaging experimental setup. A halogen lamp source produces white light. Each wavelength is rotated to a unique linear polarization angle selected to correct the wavelength-dependent focal shift of the metalens. This is done by rotation of the polarization of light incident on the WDPR and also changing the voltage of the Liquid Crystal (LC) Retarder within the WDPR. The light subsequently illuminates an object, which is imaged by the metalens, magnified using a conventional-lens relay, and recorded with a color sensitive image sensor.
       }
\end{figure}

\subsection{PSF measurements}
The experimental setup used for PSF measurement is shown in Fig.~\ref{fig:exp}(a). To capture the 3D PSF for a given input wavelength and polarization, the metasurface is translated along the optical (z-) axis, while the remainder of the imaging system is kept stationary. At each z-position, a 2D image is acquired containing a single XY-slice of the PSF. The plane that is imaged is defined by the distance of the metasurface to the objective lens. Images are acquired at z-positions ranging from the metasurface aperture plane ($z=0$) up to $z=900~{\rm \upmu m}$ beyond the aperture, in $10~{\rm \upmu m}$ steps, for a total of 90 images. These image stacks were used to characterize the metalens focal length, depth of field and PSF width at the focal plane.

The PSF measurements are fully automated. As outlined in Fig.~\ref{fig:exp}(a), four parameters are varied to capture the polarization and wavelength dependence of the three-dimensional PSF: the incident polarization, wavelength, metalens z-position and camera exposure time. To vary wavelength, we send a supercontinuum laser (NKT Photonics SuperK EXTREME EXR-20) output to an acousto optical tunable filter (AOTF; Gooch \& Housego PCAOM LFVIS). This results in monochromatic light with a bandwidth of about 5 nm in the visible spectrum. A specific wavelength is selected by driving the AOTF with a calibrated radio-frequency signal generated by a synthesizer (TPI Synthesizer TPI-1001-B). Computer-controlled variation of the RF frequency enables remote control over the optical wavelength. To control the input linear polarization from $\theta=0\mbox{--}180\degree$, we route circularly polarized light to a linear polarizer mounted on a motorized rotation mount (Thorlabs KPRM1E). The metasurface z-position is adjusted using a motorized translation stage (Thorlabs MTSA1). The translation and rotation stages receive voltage drive signals from a computer-controlled source (Thorlabs KDC101) to set appropriate position/orientation. The metalens PSF is re-imaged by an optical relay consisting of an objective lens (Nikon Plan Fluor 40x) and tube lens (200 mm focal length), passed through an output polarizer oriented at 45$\degree$, and captured on a monochrome image sensor (Basler acA1920-155um). Due to the non-uniform spectrum of the illumination source, the camera exposure time is adjusted for different wavelengths to avoid under-exposure or saturation. Motor controllers, camera settings, and the RF signal generator are controlled using custom software developed in LabView (National Instruments). Figure~\ref{fig:exp}(b) outlines the computer-controlled acquisition procedure implemented to achieve the most time-efficient and repeatable measurements.

\subsection{\label{sec:aryeh}Commercial microlens characterization}
\begin{figure}[h]
    \centering
\includegraphics[width=.9\textwidth]{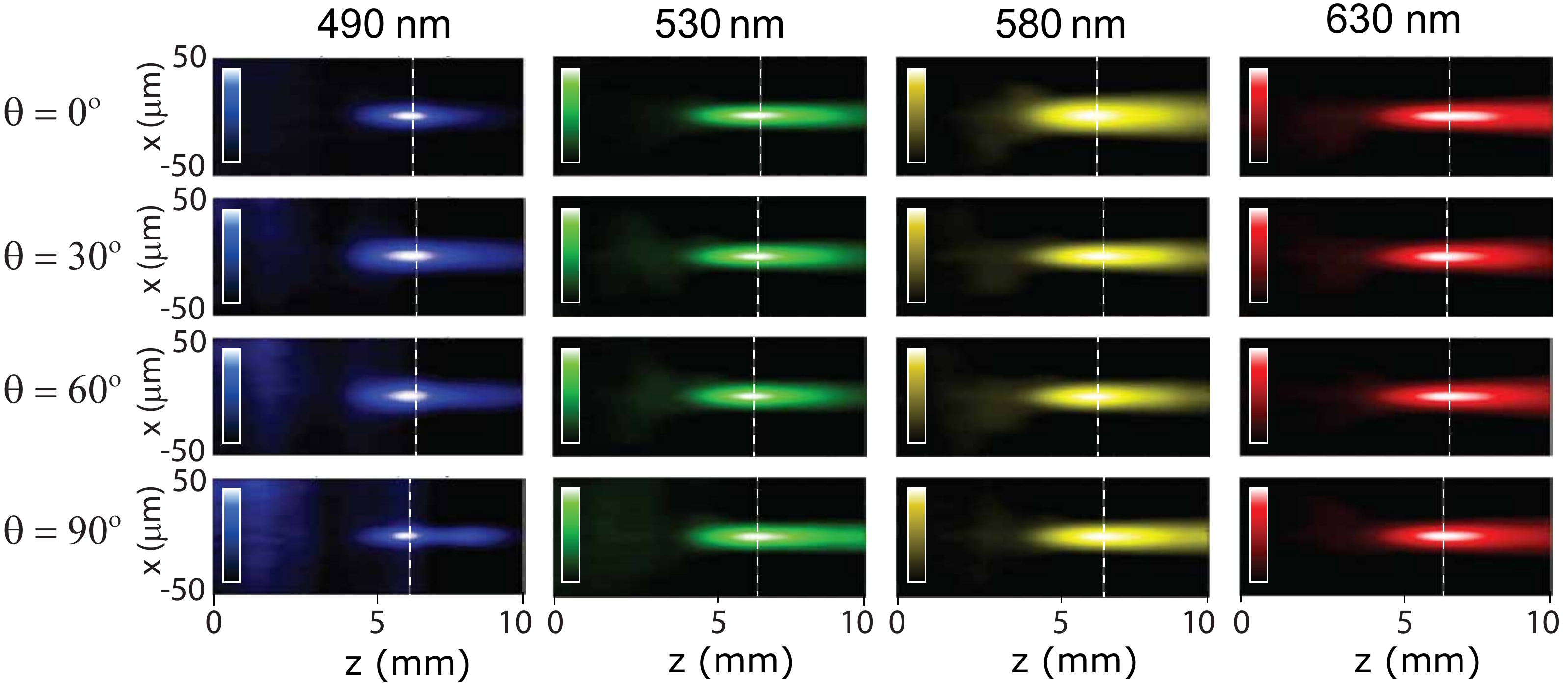}\hfill
       \caption{\label{fig:micro}
\textbf{Commercial microlens XZ slices.} XZ-slices of the measured PSF of a commercial microlens taken at representative wavelengths and polarizations. The fitted focal distances are shown as vertical white lines. 
       }
\end{figure}

As a control experiment for our metalens data acquisition and analysis techniques, we measured and analyzed the 3D PSF of a commercial microlens (Thorlabs MLA150-5C-M). We used the same setup outlined in Sec. \ref{sec:SI3} and shown in Fig.~\ref{fig:exp}(a) except we swapped the 40X for a 10X objective lens in order to make use of its longer working distance. This, in effect, lowered the magnification of our system. The results of this experiment are shown in Fig.~\ref{fig:micro} and Tab.~\hyperref[table:micro]{S1}. The observed focal length of the commercial microlens is nearly independent of both polarization and wavelength, as expected for a refractive optic. We take this as additional confirmation of the polarization-based tunability observed in the metasurface optics. 

\begin{table}[ht]
\caption{\label{table:micro} Fitted focal distances (in mm) for a commercial microlens.}
\begin{tabular}{| c | c | c | c | c |}
\hline
    & $\lambda=490~{\rm nm}$    & $\lambda=530~{\rm nm}$     & $\lambda=580~{\rm nm}$     & $\lambda=630~{\rm nm}$     \\ \hline
$\theta=0\degree$  & $6.5 \pm 0.2$ & $6.7 \pm 0.3$ & $6.5 \pm 0.3$ & $6.8 \pm 0.4$ \\ \hline
$\theta=30\degree$ & $6.6 \pm 0.3$ & $6.6 \pm 0.2$ & $6.6 \pm 0.2$ & $6.7 \pm 0.2$ \\ \hline
$\theta=60\degree$ & $6.6 \pm 0.3$ & $6.5 \pm 0.2$ & $6.5 \pm 0.2$ & $6.8 \pm 0.3$ \\ \hline
$\theta=90\degree$ & $6.4 \pm 0.6$ & $6.6 \pm 0.1$ & $6.7 \pm 0.2$ & $6.7 \pm 0.3$ \\ \hline
\end{tabular}

\end{table}

\subsection{Broadband imaging}
\label{sec:SIbroadband}

The experimental setup used for broadband imaging is shown in Fig.~\ref{fig:exp}(c). The source for this setup is a halogen lamp (Thorlabs OSL2) that outputs broadband light in the visible spectrum. To collimate this light, the source passes through a fiber bundle and is condensed by an objective lens (Nikon Plan Apo 10x). Next, the light passes through a linear polarizer on a motorized rotation mount (Thorlabs KPRM1E). This linear polarizer, in combination with the subsequent WDPR, sets the achromatic focal setting. An additional focusing lens after the WDPR is used to ensure sufficient optical intensity impinges on the resolution target(s). The resolution target used for our first broadband imaging experiment, Figs.~\ref{fig:widefield}(b-c), is a Thorlabs USAF target that is negatively patterned (Thorlabs R1L1S1N). A negatively patterned target refers to the target features being transmissive and the background being coated with chrome. The patterned surface of the target faces the metalens. The metalens forms an image of the transmissive elements which is then routed through a 12x magnification image relay comprised of an objective lens (Nikon CFI Achro LWD 40xc) and tube lens (60 mm focal length). Finally, the light passes through an output linear polarizer oriented at 45$\degree$ and is registered by a color-sensitive image sensor (Thorlabs DCC1645C). The metalens and the resolution target are both mounted on 3D translation stages for fine adjustment.

\begin{figure}[htbp]
    \centering
\includegraphics[width=.98\textwidth]{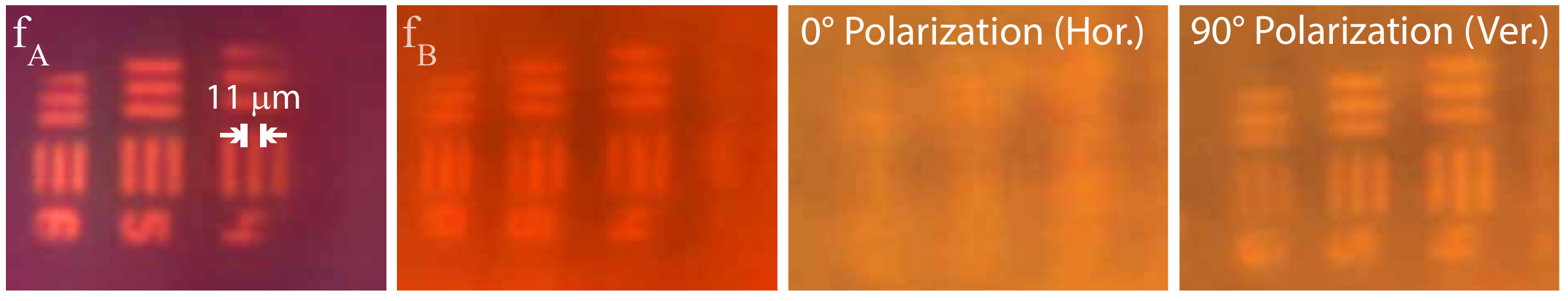}\hfill
       \caption{\label{fig:uncorrected_whitelight}
\textbf{Comparing corrected and uncorrected broadband imaging.}  All images are of USAF resolution target Group 5 Elements 4-6. (left two images) Metalens imaging of broadband light using two settings of the  focal length correction scheme described in Section \ref{sec:SIrotator}. (right two images) Metalens imaging of broadband light where all wavelengths of light have the same linear polarization, horizontal or vertical respectively.     
       }
\end{figure}

Figure~\ref{fig:uncorrected_whitelight} shows a comparison of images taken with and without the WDPR chromatic correction. Four images of the same region of the negative resolution target are shown. The first two images (from left to right) correspond to the chromatically-corrected $f_A$ and $f_B$ WDPR settings. These images are the same as those shown in Figs. \ref{fig:widefield}(b-c). The next two images are taken without the WDPR and with input polarization fixed to $\theta=0\degree$ and $\theta=90\degree$, respectively. Evidently the $f_A$ setting offers the highest image contrast and spatial resolution, owing to the higher NA and superior chromatic correction.

For focal switching experiments, Fig.~\ref{fig:widefield}(d), we used the same setup as shown in Fig.~\ref{fig:exp}(c) and imaged a ``3D'' object. The object consists of two positively-patterned resolution targets separated by 10 mm. A positively-patterned resolution target has chrome elements and a transmissive background and therefore the images have an illuminated background with shadow features. Glass microscope slides, coated with index-matching oil, were used as spacers between the targets to avoid additional reflections between air-glass interfaces. The separation distance of 10 mm was chosen due to the relatively large depth of field of the metalens (especially in the $f_B$ setting), resulting from its lower numerical aperture. A lateral offset between the imaged elements in each of the two targets is introduced to enable better visualization of the focal switching behavior. The target closest to the metalens (Thorlabs R1L3S3P) is imaged with focal length $f_A$ and the target furthest from the metalens (Thorlabs R1L3S2P) is imaged with focal length $f_B$. Switching between these foci requires rotating the linear polarizer that is on a motorized rotation mount and changing the voltage applied to the WDPR's liquid crystal. 

\begin{figure}[htbp]
    \centering
\includegraphics[width=.8\textwidth]{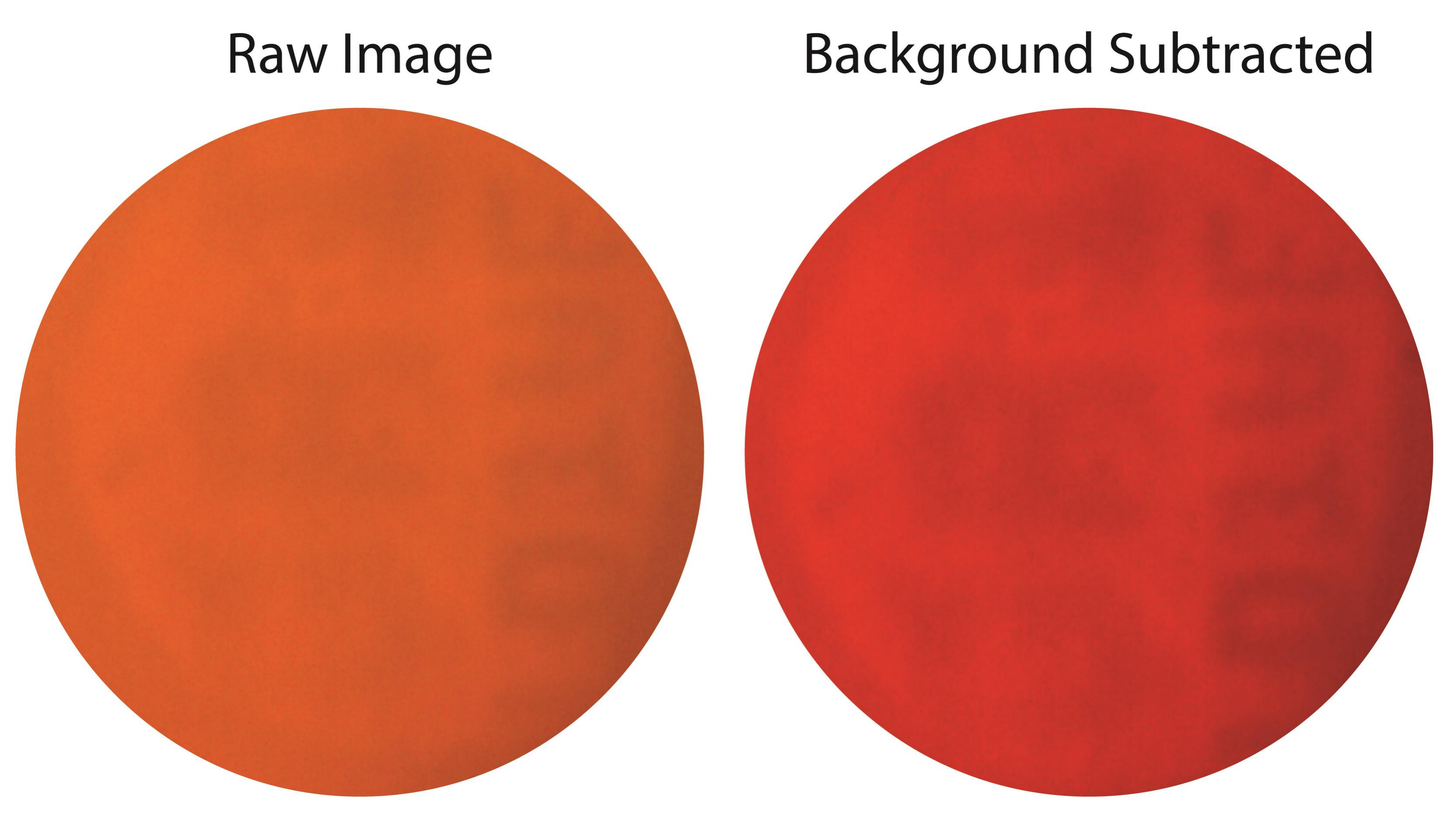}\hfill
       \caption{\label{fig:background_compare}
\textbf{Comparing raw image and background subtracted widefield images.}  Raw image (left) and background-subtracted image (right) of the 3D object taken with $f_B$ WDPR settings. The image on the right is the same as shown in Fig.~\ref{fig:widefield}(e, bottom right).  
       }
\end{figure}

As mentioned in Sec.~\label{subsec:whitelightimaging}, all color images presented in Fig.~\ref{fig:widefield} correspond to the raw RGB image sensor output, with the exception of the image of the 3D object taken with the $f_B$ setting, Fig.~\ref{fig:widefield}(e, bottom right). In this case, we applied a background subtraction to enhance the image contrast. To show the (relatively minor) effect of the background subtraction, we've included both the raw and manipulated images in Fig.~\ref{fig:background_compare}.

\subsection{Transmission and focusing efficiency}

\label{sec:SItransmission}

\begin{figure}[htbp]
    \centering
\includegraphics[width=.8\textwidth]{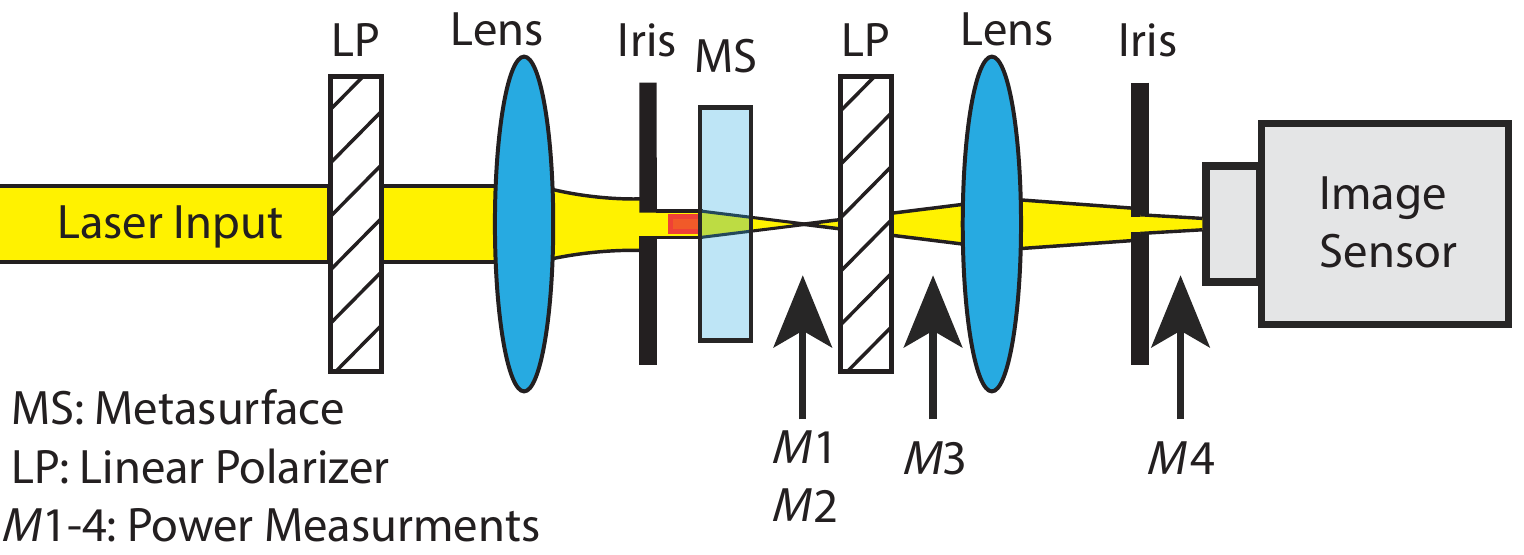}\hfill
       \caption{\label{fig:transmission}
\textbf{Experimental Setup for Transmission and Focusing Efficiency Measurements.} Experimental setup used to determine metalens efficiency metrics.
       }
\end{figure}

The transmission and focusing efficiency data shown in Fig.~\ref{fig:2}(d) was acquired using the method outlined in Fig.~\ref{fig:transmission}. The supercontinuum source light passes through the AOTF, which is set to the representative wavelength of 581 nm for this measurement. The light then passes through a linear polarizer to vary the polarization from $\theta=0\mbox{--}180\degree$. Next, the light passes through a lens and an iris to ensure that only the metasurface aperture is illuminated. Light exiting the metasurface aperture passes through the output polarizer set to $45\degree$. Finally the light passes through another lens to relay the metalens focal plane onto an image sensor. An iris is placed near the secondary image plane to block undiffracted light that does not contribute to the metalens focal region.

We measure the optical power at four different locations in the optical path, labeled $M1-4$ in Fig.~\ref{fig:transmission}. Each power is recorded as a function of input polarization angle, $\theta$. To account for background light unrelated to the illumination source, the laser is blocked and the background power, $M_{DK}$, is measured. These five measurements are used to calculate three metalens efficiency metrics: transmission efficiency without the output polarizer, transmission efficiency with the output polarizer, and focusing efficiency.

Measurement $M1$ is taken in the location just after the metalens, but with the metalens removed from the setup. This measurement records the power incident on the metalens aperture (while avoiding recording any loss from the substrate). $M2$ occurs in the same location as $M1$, but now with the metalens in place. This measurement records the power that is transmitted through the metalens aperture. These two measurements, in addition the background power, are used to calculate the transmission efficiency without the output polarizer as:

\begin{align}
    \textit{Transmission efficiency without OP} &= \frac{M2-M_{DK}}{M1-M_{DK}}.
\end{align}

Measurement $M3$ is taken directly after the output polarizer. This measurement records how much power is transmitted through both the metalens and the output polarizer. The transmission efficiency with the output polarizer is then given by:

\begin{align}
    \textit{Transmission efficiency with OP} &= \frac{M3-M_{DK}}{M1-M_{DK}}.
\end{align}

Measurement $M4$ is taken after the iris in the vicinity of the secondary metalens focal plane. The focusing efficiency of the metalens is therefore calculated as:

\begin{align}
    \textit{Focusing efficiency} &= \frac{M4-M_{DK}}{M1-M_{DK}}.
\end{align}

\subsection{Determining metalens focal length}
The metalens focal length was determined by fitting a 2D Gaussian to background-subtracted images (i. e. XY-planes),
\begin{equation}
    f(x,y)= A e^{-\frac{(x-x_0)^2}{2\sigma^2}-\frac{(y-y_0)^2}{2\sigma^2}} ,
\end{equation}
for each XY-slice in the 3D PSF. Frames where the intensity distribution was not Gaussian and/or where the fit did not converge, were dropped. The resulting fitted values of $A$ were plotted as a function of $z$. The focal length is defined as the $z$-position corresponding to the maximum of the function $A(z)$. This was determined by fitting $A(z)$ with a skewed normal distribution,
\begin{equation}
    A_z e^{-\frac{(z-z_0)^2}{2\sigma_z^2}}\text{erf}\left(-\frac{(z-z_0)\alpha}{\sqrt{2}\sigma_z}\right) ,
\end{equation}
and solving for the $z$-position of the maximum.

\bibliographystyle{apsrev4-1}

\end{document}